\documentclass[aps,prm,twocolumn,amsmath,amssymb,floatfix,superscriptaddress,citeautoscript]{revtex4-2}

\usepackage{graphicx}
\usepackage{natbib}
\usepackage{color}
\usepackage{hyperref}

\begin{document}

	\title{Importance of the semimetallic state for the quantum Hall effect in HfTe$_{5}$}
	
	\author{M.~M.\ Piva}
	\email{Mario.Piva@cpfs.mpg.de}
	\affiliation{Max Planck Institute for Chemical Physics of Solids, N\"{o}thnitzer Stra{\ss}e 40, D-01187 Dresden, Germany}

    \author{R.\ Wawrzy\'nczak}
	\affiliation{Max Planck Institute for Chemical Physics of Solids, N\"{o}thnitzer Stra{\ss}e 40, D-01187 Dresden, Germany}

    \author{Nitesh Kumar}
	\affiliation{Max Planck Institute for Chemical Physics of Solids, N\"{o}thnitzer Stra{\ss}e 40, D-01187 Dresden, Germany}
    \affiliation{S. N. Bose National Centre for Basic Sciences, Salt Lake City, Kolkata, 700 106 India}
 
	\author{L.~O.~Kutelak}
	\affiliation{Brazilian Synchrotron Light Laboratory (LNLS), Brazilian Center for Research in Energy and Materials (CNPEM), Campinas 13083-970, SP, Brazil}
	\affiliation{Max Planck Institute for Chemical Physics of Solids, N\"{o}thnitzer Stra{\ss}e  40, D-01187 Dresden, Germany}

    \author{G.~A.\ Lombardi}
	\affiliation{Brazilian Synchrotron Light Laboratory (LNLS), Brazilian Center for Research in Energy and Materials (CNPEM), Campinas 13083-970, SP, Brazil}

    \author{R.~D.~dos Reis}
	\affiliation{Brazilian Synchrotron Light Laboratory (LNLS), Brazilian Center for Research in Energy and Materials (CNPEM), Campinas 13083-970, SP, Brazil}

    \author{C. Felser}
	\affiliation{Max Planck Institute for Chemical Physics of Solids, N\"{o}thnitzer Stra{\ss}e 40, D-01187 Dresden, Germany}

	\author{M.\ Nicklas}
	\email{Michael.Nicklas@cpfs.mpg.de}
	\affiliation{Max Planck Institute for Chemical Physics of Solids, N\"{o}thnitzer Stra{\ss}e 40, D-01187 Dresden, Germany}

\date{\today}
	
\begin{abstract}
		
At ambient pressure, HfTe$_{5}$ is a material at the boundary between a weak and a strong topological phase, which can be tuned by changes in its crystalline structure or by the application of high magnetic fields. It exhibits a Lifshitz transition upon cooling, and three-dimensional (3D) quantum Hall effect (QHE) plateaus can be observed at low temperatures. Here, we have investigated the electrical transport properties of HfTe$_{5}$ under hydrostatic pressure up to 3~GPa. We find a pressure-induced crossover from a semimetallic phase at low pressures to an insulating phase at about 1.5 GPa. Our data suggest the presence of a pressure-induced Lifshitz transition at low temperatures within the insulating phase around 2~GPa. The quasi-3D QHE is confined to the low-pressure region in the semimetallic phase. This reveals the importance of the semimetallic groundstate for the emergence of the QHE in HfTe$_{5}$ and thus favors a scenario based on a low carrier density metal in the quantum limit for the observed signatures of the quasi-quantized 3D QHE.
   
\end{abstract}

\maketitle

	
The two-dimensional (2D) quantum Hall effect (QHE) occurs due to the tuning of the Fermi level through Landau levels created by the application of high magnetic fields in 2D materials, resulting in a quantized Hall conductivity with values of $\nu e^{2}/h$, where $\nu$ is the filling factor, $e$ is the electron charge, and $h$ is Planck's constant. Along with these plateaus in the Hall conductivity, the longitudinal resistivity vanishes due to conduction without scattering at the edges of the material. Remarkably, the bulk remains insulating because this effect occurs when the Fermi energy is in the gap between different Landau levels. This effect was first observed in 1980 by von Klitzing {\it et al.}\ \cite{klitzing1980new} and its discovery led to the application of topological concepts to solid-state physics \cite{thouless1982quantized}. 

Materials with non-trivial topology in momentum space or real space have great potential for applications in many fields, from catalysis and solar cells to spintronics and even high-energy physics \cite{kumar2020topological}. In particular, topological insulators possess an insulating bulk and a metallic surface due to the presence of metallic surface states similar to the state responsible for the QHE. These materials can be classified as weak topological insulators (WTIs) or strong (STIs) topological insulators, depending on whether the surface states are present only on the side surfaces of the material or on all surfaces \cite{fu2007topological,hasan2011three}. 

The 2D QHE has been extensively studied and found in graphene, for example \cite{zhang2005experimental,jiang2007quantum}. However, its three-dimensional (3D) counterpart was proposed theoretically in 1987 \cite{halperin1987possible}, but it is still difficult to realize and observe experimentally. One way to achieve a 3D QHE analog state is to stack 2D quantum Hall layers in a highly anisotropic environment \cite{avron1983homotopy,stormer1986quantization,gooth2023quantum}. However, this leads to physics similar to that observed in the 2D QHE and is referred to as quasi-2D QHE. This effect has been observed in Bechgaard salts \cite{hannahs1989quantum,cooper1989quantized}, $\eta$-Mo$_{4}$O$_{11}$ \cite{hill1998bulk}, graphite \cite{kopelevich2003reentrant,kopelevich2009searching,yaguchi2009high}, Bi$_{2}$Se$_{3}$ \cite{cao2012quantized}, EuMnBi$_{2}$ \cite{masuda2016quantum} and BaMnSb$_{2}$ \cite{sakai2020bulk,liu2017superconductivity}.

Recently, the 3D QHE has been observed in the transition metal pentatellurides ZrTe$_{5}$ and HfTe$_{5}$ \cite{tang2019three,wang2020approaching,galeski2020unconventional,galeski2021origin}. Both materials crystallize in an orthorhombic $Cmcm$ structure with TMTe$_{5}$ (TM = transition metal) layers stacked along the $b$ axis \cite{furuseth1973crystal,fjellvaag1986structural}. The groundstates of these compounds depend strongly on the volume of the unit cell \cite{fan2017transition,facio2023engineering}, which can be influenced by the growth conditions of the crystals. They can be considered as Dirac semimetals closely related to an STI or WTI phase \cite{galeski2020unconventional,galeski2021origin,gooth2023quantum}. Moreover, upon cooling, a Lifshitz transition can be observed in both, which is a change in the Fermi surface topology \cite{lifshitz1960anomalies}, changing the dominant carrier type from holes to electrons and driving the materials into a semiconductor phase and back to a semimetallic phase at low temperatures \cite{zhang2017electronic,zhang2017temperature}. The origin of the QHE is still under debate. Previous reports found that the quantized Hall effect is caused by Fermi surface instabilities such as a charge density wave (CDW) \cite{tang2019three,wang2020approaching,qin2020theory}. On the other hand, several reports claim that quasi-quantized Hall plateaus and longitudinal resistivity minima exist in the quantum limit of 3D metals with low carrier density and closed Fermi surfaces \cite{galeski2021origin,gooth2023quantum}. The main difference between these two scenarios is the need for a gapped groundstate (in the first scenario) or a metallic groundstate (in the second scenario).

One way to distinguish between these two scenarios is to study the evolution of the QHE as the material groundstate is tuned from metal to insulator. Electronic band structure calculations revealed that the activation energy of HfTe$_{5}$ increases upon compression \cite{fan2017transition}, suggesting that it may be possible to tune this material from its Dirac semimetallic state at ambient pressure to an insulating state under compression. In contrast, previous electrical resistivity measurements under pressure have found a metallic state with increasing pressure \cite{qi2016pressure,liu2017superconductivity}.  A possible explanation could be non-hydrostatic pressure conditions, since the previous pressure experiments were performed with either no \cite{qi2016pressure} or a solid \cite{liu2017superconductivity} pressure-transmitting medium. This possibility is further supported by recent  measurements under uniaxial stress and calculations which found that a metallic groundstate is favored for uniaxial compressive stress \cite{liu2023controllable,jo2023effects}, i.e., non-hydrostatic pressure conditions.

In this Letter we focus on the effects of hydrostatic pressure on the electronic properties of HfTe$_{5}$. We show that HfTe$_{5}$ can be tuned from its Dirac semimetallic state at ambient pressure to a gapped groundstate under hydrostatic compression.  We find the signature of the quasi-quantized 3D QHE only at low pressures in the semimetallic region of the temperature -- pressure phase diagram. In addition, we find a change in the type of dominant charge carriers at low temperatures and high pressures, indicating that HfTe$_{5}$ exhibits another pressure-induced Lifshitz transition.  Our results emphasize the importance of hydrostatic conditions in pressure experiments and pinpoint the origin of the 3D QHE in HfTe$_{5}$ as a feature present in the quantum limit of metals with low carrier density and closed Fermi surfaces. 

\begin{figure}[!t]
    \includegraphics[width=0.95\linewidth]{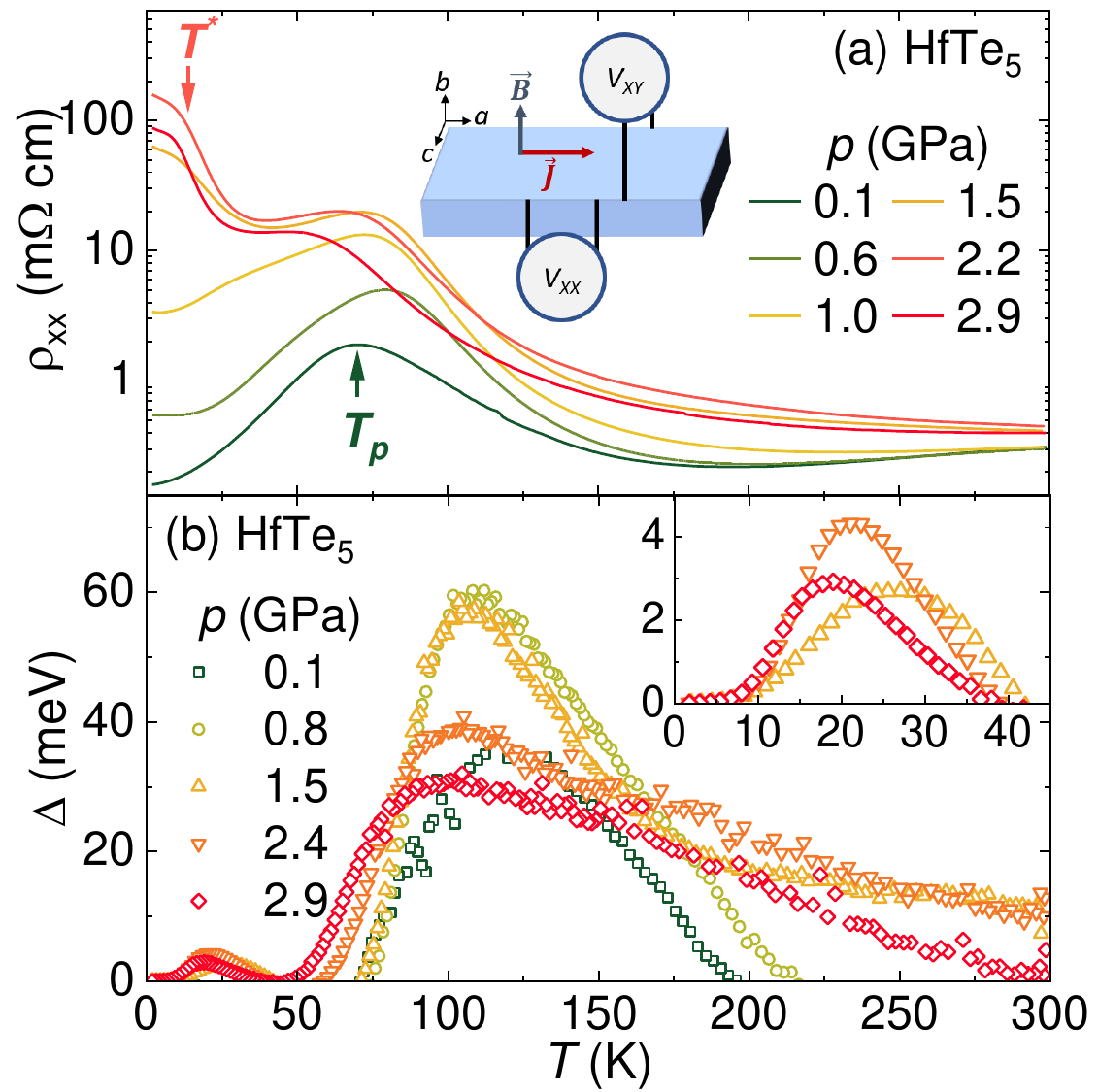}
	\caption{(a) Longitudinal resistivity ($\rho_{xx}$) as a function of temperature at various pressures. The inset presents a schematic drawing of the sample used in the experiments. (b) Activation energy ($\Delta$) as a function of temperature at selected pressures. The inset shows a detailed view of the low-temperature region.}
	\label{RxT}\label{GAP3}
\end{figure}

	
Single crystals of HfTe$_{5}$ were grown by chemical vapor transport method. Stoichiometric amounts of Hf and Te were sealed in a quartz ampoule with iodine and placed in a two-zone furnace for over a month. Electrical resistivity measurements were performed with a four-probe configuration as shown in the inset of Fig.~\ref{RxT} (a) using a commercial Physical Properties Measurement System (Quantum Design) along with an AC resistance bridge (Linear Research).  Pressures up to 2.9~GPa were generated utilizing a self-clamped piston-cylinder-type pressure cell. Silicon oil served as the pressure transmitting medium in order to avoid non-hydrostatic pressure conditions in the investigated pressure range \cite{nicklas2015pressure} and lead was the pressure gauge.


Figure~\ref{RxT}(a) presents the temperature dependence of the longitudinal resistivity ($\rho_{xx}$)  at selected pressures. At 0.1~GPa, close to ambient pressure, a large peak centered around 70~K is observed upon cooling due to the presence of a temperature-induced Lifshitz transition, in agreement with previous reports \cite{fuller1983pressure,zhang2017temperature,wang2020approaching,galeski2020unconventional}. The chemical potential of HfTe$_{5}$ is strongly temperature dependent. At room temperature the dominant carriers are hole-like; upon cooling the system becomes a semiconductor and then a semimetal again, but with dominant electron-like charge carriers below the Lifshitz transition \cite{zhang2017temperature}. 

\begin{figure}[!t]
	\includegraphics[width=0.93\linewidth]{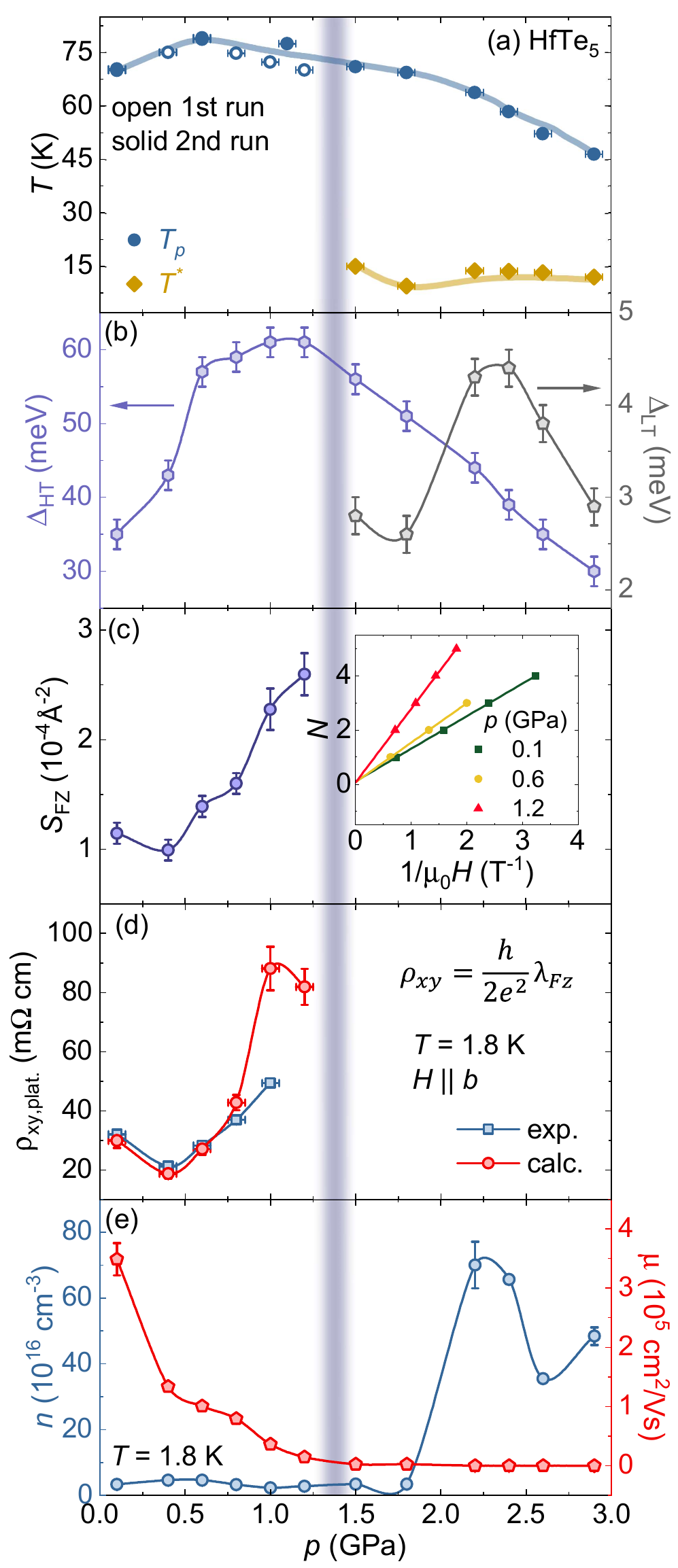}
	\caption{(a) Temperature-pressure phase diagram of HfTe$_{5}$. (b) Maximum values of the high-temperature gap ($\Delta_{HT}$) (left axis) and low-temperature gap ($\Delta_{LT}$)  (right axis) as a function of pressure. (c) Cross sectional area of the Fermi surface ($S_{FZ}$) as a function of pressure. The inset presents a Landau fan diagram at selected pressures. (d) Calculated $\rho_{xy{\rm,plat.}}^{\rm cal.}$ and measured Hall resistivity $\rho_{xy{\rm,plat.}}^{\rm exp.}$ at the last plateau of the QHE. (e) Carrier density ($n$) and mobility ($\mu$) as a function of pressure. The jump in $n(p)$ is accompanied by a change in carrier type from electron-like to hole-like.}
	\label{pdep}
\end{figure}

Application of external pressure first shifts the position of the peak ($T_{p}$) to higher temperatures, reaching a maximum of 79~K at 0.6~GPa, further increasing pressure suppresses $T_{p}$ down to 47~K at 2.9~GPa (see Fig.~\ref{pdep}(a)), in good agreement with previous reports \cite{fuller1983pressure,liu2017superconductivity} and similar to observations in the sister compound ZrTe$_{5}$ \cite{zhou2016pressure}. 
However, the strongest effect of pressure is seen in the longitudinal resistivity at  low temperatures (see Fig.~\ref{RxT}(a) ). The metallic behavior below $T_{p}$ first becomes weaker with increasing pressure, and above 1.5~GPa $\rho_{xx}(T)$ starts to increase exponentially toward low temperatures, indicating the opening of a gap below $\approx30$~K that has not been reported before, possibly due to non-hydrostatic pressure conditions in the previous experiments. We will return to this latter point below. Note that around 15~K $\rho_{xx}(T)$ tends to saturate toward low temperatures, and a temperature $T^{*}$ can be defined as an inflection point in $\rho_{xx}(T)$. It is almost pressure independent up to 2.9~GPa. The origin of this behavior is still unknown and could be related to one of the following scenarios: a conduction dominated by topological surface states \cite{liu2023controllable}, or the presence of impurity bands \cite{seeger2013semiconductor} or a side conduction band, both of which can contribute to electronic transport \cite{zhang2017temperature}.

To gain further insight into the electronic structure of HfTe$_{5}$, we investigated the evolution of the activation energy ($\Delta$) as a function of temperature and pressure. The gap energy is usually obtained as the linear slope of an Arrhenius plot $\ln(\rho_{xx}) = \ln(\rho_{0}) + \Delta / T$. However, in HfTe$_{5}$ the activation energy shows a strong evolution as a function of temperature \cite{zhang2017temperature}, so it is more appropriate to analyze the temperature derivative of the Arrhenius plot $\ln(\rho_{xx}) = \ln(\rho_{0}) + \Delta(T) / T$, which yields $\Delta(T)$, as shown in Fig.~\ref{GAP3}(b).  At low pressures we observe only the opening of a gap in the high-temperature region ($T\gtrsim50$~K) and metallic behavior below 50~K, but at 1.5~GPa and above we find the opening of a second gap, or the reopening of the high-temperature gap, at low temperatures ($T\lesssim50$~K). We will refer to the maximum size of the high-temperature gap as $\Delta_{HT}$ and the low-temperature gap as $\Delta_{LT}$.

At 0.1~GPa near ambient pressure, the high-temperature gap opens at 200~K and reaches a maximum size of  $\Delta_{HT}=35$~meV at around 120~K, in agreement with previous reports \cite{zhang2017temperature}. With increasing pressure, $\Delta_{HT}$ increases to 61~meV at 1.1~GPa (see Fig.~\ref{pdep}(b)), in agreement with electronic band calculations upon unit cell compression \cite{fan2017transition} and with an early study of the electrical resistivity HfTe$_{5}$ under pressure \cite{fuller1983pressure}. Above 1.2~GPa $\Delta_{HT}$ decreases with further increasing pressure, but the gap opens  already at higher temperatures and beyond 1.5~GPa already above room temperature. Furthermore, around 1.5~GPa the temperature dependence of the resistivity also indicates a gap at low temperatures, $\Delta_{LT}$ reaching a value of 2.8~meV at about 25~K at 1.5~GPa. By increasing the pressure  $\Delta_{LT}$ increases strongly to 4.4~meV at 2.4~GPa and then decreases again to 2.9~meV at 2.9~GPa (see Fig.~\ref{pdep}(b)). 

A possible scenario for the existence of two different gaps in the high-temperature and low-temperature range is the occurrence of a topological phase transition from a WTI to a STI. This transition leads to the closing and reopening of the gap and it was theoretically proposed for HfTe$_{5}$ \cite{weng2014transition} upon cooling, but has not yet been observed experimentally \cite{zhang2017temperature}. Similar topological transitions in HfTe$_{5}$ have been observed as a function of uniaxial strain \cite{liu2023controllable} or magnetic fields \cite{wu2023topological}. Therefore, it is quite feasible that the application of external pressure tunes the Fermi energy of HfTe$_{5}$ to a level at which a topological phase transition can be induced by cooling.

We note that other reports have found metallic behavior with increasing applied pressure. In these experiments, a solid \cite{liu2017superconductivity} transmitting medium or even no transmitting medium \cite{qi2016pressure} was used, both of which are expected to induce a large pressure gradient in the sample chamber. These conditions show similarities to uniaxial stress experiments which favored a metallic state \cite{liu2023controllable,jo2023effects} and therefore the contrasting results of the aforementioned reports compared to our Letter are not surprising.  This demonstrates the importance of hydrostatic pressure conditions for measurements on HfTe$_{5}$ crystals.

\begin{figure}[!t]
	\includegraphics[width=0.95\linewidth]{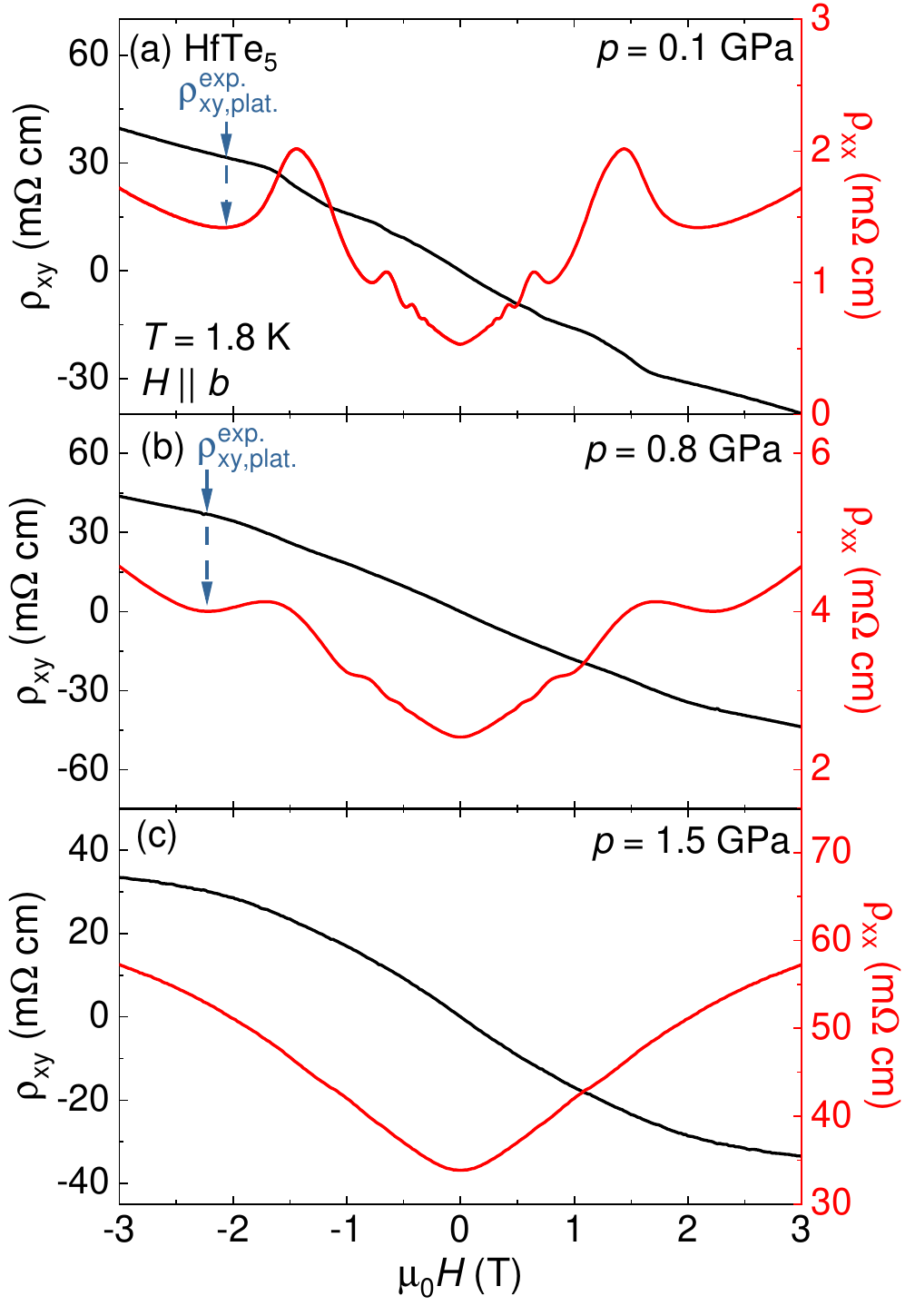}
	\caption{Longitudinal ($\rho_{xx}$) and Hall resistivity ($\rho_{xy}$) as a function of magnetic field at 1.8~K and at (a) 0.1, (b) 0.8, and (c) 1.5~GPa. The magnetic field was applied parallel to the $b$ axis in both measurements. The arrows mark the last plateau.}
	\label{QHE}
\end{figure}

At ambient pressure, the magnetoresistance (${\rm MR}$) $\frac{\rho_{xx}(H)-\rho_{xx}(0)}{\rho_{xx}(0)}$ of the investigated HfTe$_{5}$ sample shows an extremely high value of 1650\% at 9~T, indicating the high quality of the crystal (see the Supplemental Material - SM  for details \cite{SM}).  At small magnetic fields $\rho_{xx}(H)$ shows the well-known quantum oscillations (QOs) and the quantum limit is reached at 1.4~T (see Fig.~\ref{QHE}(a)). The application of external pressure suppresses the oscillations and the magnitude of the MR. At 1.5~GPa and above, where a second gap opens at low temperatures, no QOs are visible in $\rho_{xx}(H)$ (see Fig.~\ref{QHE}(c)). In addition, at low temperatures we find a change from electron-like charge carriers  to hole-like charge carriers in the Hall resistivity $\rho_{xy}$ around 2.2~GPa (see the SM \cite{SM}). This suggests the possibility of a pressure-induced Lifshitz transition. Since previous experiments were performed under less homogeneous pressure conditions than in our experiment, it may be sensitive to the pressure homogeneity. We note that our X-ray powder diffraction data do not show any structural phase transition at room temperature (see the SM \cite{SM}). In the following, we discuss the pressure evolution of the quasi-quantized 3D QHE and its relationship to the pressure-induced changes in the electronic properties.


To get more information about the changes in the Fermi surface, we analyze the QO (see the SM for details \cite{SM}). The frequency of the oscillations $B_{Fz}$ was extracted from the slope of the Landau fan diagrams, as shown in the inset of Fig.~\ref{pdep}(c). At 0.1~GPa the extracted frequency of 1.2~T is in good agreement with previous reports \cite{galeski2020unconventional,wang2020approaching,wu2023topological}, the application of external pressure enhances the oscillation frequency to 2.7~T at 1.2~GPa. At higher pressures, the data no longer show QO. The cross-sectional area of the Fermi surface $S_{Fz} = B_{Fz} \times \left( \frac{2 \pi e}{\hbar} \right)$, where $\hbar$ is the reduced Planck constant, first  remains constant with increasing pressure before it starts to increase above 0.5~GPa. The dependence of the Fermi surface area on pressure is depicted in Fig.~\ref{pdep}(c). 

Looking more closely at the low-field region, evidence for the 3D integer QHE in HfTe$_{5}$ can be seen in Fig.~\ref{QHE}(a). Plateaus in the Hall resistivity $\rho_{xy}$ occur at the same fields as the minima in the longitudinal resistivity $\rho_{xx}$.  The experimental $\rho_{xy}$ of the last plateau is close to $\rho_{xy{\rm,plat.}}^{\rm cal.}=\frac{h}{2e^{2}} \lambda_{Fz}$, where $h$ is Planck's constant, $e$ is the electron charge, and $\lambda_{Fz}$ is the Fermi wavelength along the $z$ direction. However, the longitudinal resistivity does not vanish completely as it should for a fully quantized 3D QHE. With increasing pressure the features in $\rho_{xy}$ become less pronounced. The plateaus in $\rho_{xy}$ can only be identified in the metallic regime, which is indicated by the temperature dependence of the resistivity $\rho_{xx}$ [see, e.g.,\ Figs~\ref{QHE}(a) and \ref{QHE}(b)]. At 1.5~GPa, the temperature dependence of the resistivity shown in Fig.~\ref{RxT}(a) indicates the opening of a low-temperature gap, i.e., an insulating behavior, and no signs of a 3D QHE are visible in $\rho_{xx}(H)$ and $\rho_{xy}(H)$ (Fig.~\ref{QHE}(c)).


Two possible scenarios are currently being discussed to explain a quasi-3D QHE in HfTe$_5$. The first considers the material as a stack of 2D layers that generate a Fermi surface instability, such as a CDW, which opens a gap giving rise to a quantization of the quantum Hall effect proportional to the Fermi wavelength along the $z$ direction $\lambda_{Fz}$  \cite{tang2019three,wang2020approaching}. The finite longitudinal resistivity $\rho_{xx}$ is then explained by carriers from other Fermi surface pockets contributing to the electrical transport \cite{wang2020approaching}. The second scenario achieves a quantization by considering a bulk 3D metallic material, rather than a stack of 2D layers, with a closed Fermi surface and low carrier density \cite{galeski2021origin,gooth2023quantum}. When the system enters the quantum limit, the only allowed dispersion is parallel to the applied magnetic field, resulting in metallic Landau bands that give rise to the finite longitudinal resistivity. In both cases, the value of $\rho_{xy}$ at the last plateau is given by $\frac{h}{2e^{2}} \lambda_{Fz}$ \cite{gooth2023quantum}.

At low pressures, the Hall resistivity at the last plateau agrees quite well with the prediction in both scenarios  $\rho_{xy}=\frac{h}{2e^{2}}\lambda_{Fz}$. But at 0.8~GPa the calculated value starts to become larger than the experimental one (see Fig.~\ref{QHE}(d)). In this pressure range, HfTe$_{5}$ shows no fundamental changes in its electronic properties at low temperature. It continues to exhibit metallic behavior with low carrier density.  We do not find any additional QO frequencies that could indicate abrupt changes in the Fermi surface. Moreover, the cross-sectional area of the Fermi surface $S_{FZ}$ increases monotonically with pressure; i.e.,\ we can assume that the closed 3D anisotropic Fermi pocket identified at ambient pressure, which is a prerequisite for the model assuming a bulk 3D metal \cite{galeski2020unconventional}, does not change significantly in this pressure range.  The characteristic changes in the low-temperature behavior occur at higher pressures. The low-temperature gap opens only above 1.5~GPa and a change in the dominant type of charge carrier takes place above 2~GPa. It is therefore surprising that $\rho_{xy{\rm,plat.}}^{\rm exp.}$ deviates from $\rho_{xy{\rm,plat.}}^{\rm cal.}$. The amplitude of the QO is strongly suppressed with pressure, which is caused by the strong decrease of the charge carrier mobility (see Fig.~\ref{QHE}(e)).  However, the increased uncertainty in determining the position of the last plateau does not explain the observed discrepancy. Our results support the scenario that the quasi-3D QHE in HfTe$_5$ originates from a bulk 3D metallic material with a closed Fermi surface and low carrier density, and make a scenario based on a stack of 2D layers generating a Fermi surface instability less likely.


In conclusion, we have demonstrated that the signatures of the quasi-quantized 3D QHE  in pressurized HfTe$_{5}$ are confined to the semimetallic state. This supports a scenario based on a bulk 3D metal with low carrier density in the quantum limit instead of a scenario based on stacked 2D layers and Fermi surface instabilities for the origin of the quasi-quantized 3D QHE in HfTe$_{5}$. Furthermore, we find a change from electron-like to hole-like carriers at low temperatures around 2.2~GPa, pointing to the existence of a pressure-induced Lifshitz transition.\\

\section*{DATA AVAILABILITY}
Data that underpin the ﬁndings of this Letter are available at Edmond – the open research data repository of the Max Planck Society at \cite{EDMOND}.

\begin{acknowledgments}
We acknowledge fruitful discussions with J. Gooth. This project has received funding from the European Union’s Horizon 2020 research and innovation programme under the Marie Sk\l{}odowska-Curie grant agreement No 101019024. R.D.dR. and L.O.K. acknowledges financial support from the Brazilian agencie FAPESP (Grants: 2018/00823-0, 2021/02314-9 and 2022/05447-2). R.D.dR. and G.A.L. has received financial support from the Max Planck Society under the auspices of the Max Planck Partner Group R. D. dos Reis of the MPI for Chemical Physics of Solids, Dresden, Germany. This research used facilities of the Brazilian Synchrotron Light Laboratory (LNLS), part of the Brazilian Center for Research in Energy and Materials (CNPEM), a private non-profit organization under the supervision of the Brazilian Ministry for Science, Technology, and Innovations (MCTI). The EMA beamline and LCTE staff are acknowledged for the assistance during the experiments 20220918. C.F. was financially supported by Deutsche Forschungsgemeinschaft (DFG) under SFB
1143 (Project No. 247310070) and W\"{u}rzburg-Dresden Cluster of Excellence on Complexity and
Topology in Quantum Matter—ct.qmat (EXC 2147, project no. 39085490).\\
\end{acknowledgments}

\bibliography{HfTe5}

\begin{thebibliography}{44}%
\makeatletter
\providecommand \@ifxundefined [1]{%
 \@ifx{#1\undefined}
}%
\providecommand \@ifnum [1]{%
 \ifnum #1\expandafter \@firstoftwo
 \else \expandafter \@secondoftwo
 \fi
}%
\providecommand \@ifx [1]{%
 \ifx #1\expandafter \@firstoftwo
 \else \expandafter \@secondoftwo
 \fi
}%
\providecommand \natexlab [1]{#1}%
\providecommand \enquote  [1]{``#1''}%
\providecommand \bibnamefont  [1]{#1}%
\providecommand \bibfnamefont [1]{#1}%
\providecommand \citenamefont [1]{#1}%
\providecommand \href@noop [0]{\@secondoftwo}%
\providecommand \href [0]{\begingroup \@sanitize@url \@href}%
\providecommand \@href[1]{\@@startlink{#1}\@@href}%
\providecommand \@@href[1]{\endgroup#1\@@endlink}%
\providecommand \@sanitize@url [0]{\catcode `\\12\catcode `\$12\catcode
  `\&12\catcode `\#12\catcode `\^12\catcode `\_12\catcode `\%12\relax}%
\providecommand \@@startlink[1]{}%
\providecommand \@@endlink[0]{}%
\providecommand \url  [0]{\begingroup\@sanitize@url \@url }%
\providecommand \@url [1]{\endgroup\@href {#1}{\urlprefix }}%
\providecommand \urlprefix  [0]{URL }%
\providecommand \Eprint [0]{\href }%
\providecommand \doibase [0]{https://doi.org/}%
\providecommand \selectlanguage [0]{\@gobble}%
\providecommand \bibinfo  [0]{\@secondoftwo}%
\providecommand \bibfield  [0]{\@secondoftwo}%
\providecommand \translation [1]{[#1]}%
\providecommand \BibitemOpen [0]{}%
\providecommand \bibitemStop [0]{}%
\providecommand \bibitemNoStop [0]{.\EOS\space}%
\providecommand \EOS [0]{\spacefactor3000\relax}%
\providecommand \BibitemShut  [1]{\csname bibitem#1\endcsname}%
\let\auto@bib@innerbib\@empty
\bibitem [{\citenamefont {Klitzing}\ \emph {et~al.}(1980)\citenamefont
  {Klitzing}, \citenamefont {Dorda},\ and\ \citenamefont
  {Pepper}}]{klitzing1980new}%
  \BibitemOpen
  \bibfield  {author} {\bibinfo {author} {\bibfnamefont {K.~v.}\ \bibnamefont
  {Klitzing}}, \bibinfo {author} {\bibfnamefont {G.}~\bibnamefont {Dorda}},\
  and\ \bibinfo {author} {\bibfnamefont {M.}~\bibnamefont {Pepper}},\
  }\bibfield  {title} {\bibinfo {title} {{New method for high-accuracy
  determination of the fine-structure constant based on quantized Hall
  resistance}},\ }\href {https://doi.org/10.1103/PhysRevLett.45.494} {\bibfield
   {journal} {\bibinfo  {journal} {Phys. Rev. Lett.}\ }\textbf {\bibinfo
  {volume} {45}},\ \bibinfo {pages} {494} (\bibinfo {year} {1980})}\BibitemShut
  {NoStop}%
\bibitem [{\citenamefont {Thouless}\ \emph {et~al.}(1982)\citenamefont
  {Thouless}, \citenamefont {Kohmoto}, \citenamefont {Nightingale},\ and\
  \citenamefont {den Nijs}}]{thouless1982quantized}%
  \BibitemOpen
  \bibfield  {author} {\bibinfo {author} {\bibfnamefont {D.~J.}\ \bibnamefont
  {Thouless}}, \bibinfo {author} {\bibfnamefont {M.}~\bibnamefont {Kohmoto}},
  \bibinfo {author} {\bibfnamefont {M.~P.}\ \bibnamefont {Nightingale}},\ and\
  \bibinfo {author} {\bibfnamefont {M.}~\bibnamefont {den Nijs}},\ }\bibfield
  {title} {\bibinfo {title} {{Quantized Hall conductance in a two-dimensional
  periodic potential}},\ }\href {https://doi.org/10.1103/PhysRevLett.49.405}
  {\bibfield  {journal} {\bibinfo  {journal} {Phys. Rev. Lett.}\ }\textbf
  {\bibinfo {volume} {49}},\ \bibinfo {pages} {405} (\bibinfo {year}
  {1982})}\BibitemShut {NoStop}%
\bibitem [{\citenamefont {Kumar}\ \emph {et~al.}(2020)\citenamefont {Kumar},
  \citenamefont {Guin}, \citenamefont {Manna}, \citenamefont {Shekhar},\ and\
  \citenamefont {Felser}}]{kumar2020topological}%
  \BibitemOpen
  \bibfield  {author} {\bibinfo {author} {\bibfnamefont {N.}~\bibnamefont
  {Kumar}}, \bibinfo {author} {\bibfnamefont {S.~N.}\ \bibnamefont {Guin}},
  \bibinfo {author} {\bibfnamefont {K.}~\bibnamefont {Manna}}, \bibinfo
  {author} {\bibfnamefont {C.}~\bibnamefont {Shekhar}},\ and\ \bibinfo {author}
  {\bibfnamefont {C.}~\bibnamefont {Felser}},\ }\bibfield  {title} {\bibinfo
  {title} {Topological quantum materials from the viewpoint of chemistry},\
  }\href {https://doi.org/10.1021/acs.chemrev.0c00732} {\bibfield  {journal}
  {\bibinfo  {journal} {Chem. Rev.}\ }\textbf {\bibinfo {volume} {121}},\
  \bibinfo {pages} {2780} (\bibinfo {year} {2020})}\BibitemShut {NoStop}%
\bibitem [{\citenamefont {Fu}\ \emph {et~al.}(2007)\citenamefont {Fu},
  \citenamefont {Kane},\ and\ \citenamefont {Mele}}]{fu2007topological}%
  \BibitemOpen
  \bibfield  {author} {\bibinfo {author} {\bibfnamefont {L.}~\bibnamefont
  {Fu}}, \bibinfo {author} {\bibfnamefont {C.~L.}\ \bibnamefont {Kane}},\ and\
  \bibinfo {author} {\bibfnamefont {E.~J.}\ \bibnamefont {Mele}},\ }\bibfield
  {title} {\bibinfo {title} {Topological insulators in three dimensions},\
  }\href {https://doi.org/10.1103/PhysRevLett.98.106803} {\bibfield  {journal}
  {\bibinfo  {journal} {Phys. Rev. Lett.}\ }\textbf {\bibinfo {volume} {98}},\
  \bibinfo {pages} {106803} (\bibinfo {year} {2007})}\BibitemShut {NoStop}%
\bibitem [{\citenamefont {Hasan}\ and\ \citenamefont
  {Moore}(2011)}]{hasan2011three}%
  \BibitemOpen
  \bibfield  {author} {\bibinfo {author} {\bibfnamefont {M.~Z.}\ \bibnamefont
  {Hasan}}\ and\ \bibinfo {author} {\bibfnamefont {J.~E.}\ \bibnamefont
  {Moore}},\ }\bibfield  {title} {\bibinfo {title} {Three-dimensional
  topological insulators},\ }\href
  {https://doi.org/10.1146/annurev-conmatphys-062910-140432} {\bibfield
  {journal} {\bibinfo  {journal} {Annu. Rev. Condens. Matter Phys.}\ }\textbf
  {\bibinfo {volume} {2}},\ \bibinfo {pages} {55} (\bibinfo {year}
  {2011})}\BibitemShut {NoStop}%
\bibitem [{\citenamefont {Zhang}\ \emph {et~al.}(2005)\citenamefont {Zhang},
  \citenamefont {Tan}, \citenamefont {Stormer},\ and\ \citenamefont
  {Kim}}]{zhang2005experimental}%
  \BibitemOpen
  \bibfield  {author} {\bibinfo {author} {\bibfnamefont {Y.}~\bibnamefont
  {Zhang}}, \bibinfo {author} {\bibfnamefont {Y.-W.}\ \bibnamefont {Tan}},
  \bibinfo {author} {\bibfnamefont {H.~L.}\ \bibnamefont {Stormer}},\ and\
  \bibinfo {author} {\bibfnamefont {P.}~\bibnamefont {Kim}},\ }\bibfield
  {title} {\bibinfo {title} {{Experimental observation of the quantum Hall
  effect and Berry's phase in graphene}},\ }\href
  {https://doi.org/10.1038/nature04235} {\bibfield  {journal} {\bibinfo
  {journal} {Nature}\ }\textbf {\bibinfo {volume} {438}},\ \bibinfo {pages}
  {201} (\bibinfo {year} {2005})}\BibitemShut {NoStop}%
\bibitem [{\citenamefont {Jiang}\ \emph {et~al.}(2007)\citenamefont {Jiang},
  \citenamefont {Zhang}, \citenamefont {Tan}, \citenamefont {Stormer},\ and\
  \citenamefont {Kim}}]{jiang2007quantum}%
  \BibitemOpen
  \bibfield  {author} {\bibinfo {author} {\bibfnamefont {Z.}~\bibnamefont
  {Jiang}}, \bibinfo {author} {\bibfnamefont {Y.}~\bibnamefont {Zhang}},
  \bibinfo {author} {\bibfnamefont {Y.-W.}\ \bibnamefont {Tan}}, \bibinfo
  {author} {\bibfnamefont {H.}~\bibnamefont {Stormer}},\ and\ \bibinfo {author}
  {\bibfnamefont {P.}~\bibnamefont {Kim}},\ }\bibfield  {title} {\bibinfo
  {title} {{Quantum Hall effect in graphene}},\ }\href
  {https://doi.org/10.1016/j.ssc.2007.02.046} {\bibfield  {journal} {\bibinfo
  {journal} {Solid State Commun.}\ }\textbf {\bibinfo {volume} {143}},\
  \bibinfo {pages} {14} (\bibinfo {year} {2007})}\BibitemShut {NoStop}%
\bibitem [{\citenamefont {Halperin}(1987)}]{halperin1987possible}%
  \BibitemOpen
  \bibfield  {author} {\bibinfo {author} {\bibfnamefont {B.~I.}\ \bibnamefont
  {Halperin}},\ }\bibfield  {title} {\bibinfo {title} {Possible states for a
  three-dimensional electron gas in a strong magnetic field},\ }\href
  {https://doi.org/10.7567/JJAPS.26S3.1913} {\bibfield  {journal} {\bibinfo
  {journal} {Jpn. J. Appl. Phys.}\ }\textbf {\bibinfo {volume} {26}},\ \bibinfo
  {pages} {1913} (\bibinfo {year} {1987})}\BibitemShut {NoStop}%
\bibitem [{\citenamefont {Avron}\ \emph {et~al.}(1983)\citenamefont {Avron},
  \citenamefont {Seiler},\ and\ \citenamefont {Simon}}]{avron1983homotopy}%
  \BibitemOpen
  \bibfield  {author} {\bibinfo {author} {\bibfnamefont {J.~E.}\ \bibnamefont
  {Avron}}, \bibinfo {author} {\bibfnamefont {R.}~\bibnamefont {Seiler}},\ and\
  \bibinfo {author} {\bibfnamefont {B.}~\bibnamefont {Simon}},\ }\bibfield
  {title} {\bibinfo {title} {Homotopy and quantization in condensed matter
  physics},\ }\href {https://doi.org/10.1103/PhysRevLett.51.51} {\bibfield
  {journal} {\bibinfo  {journal} {Phys. Rev. Lett.}\ }\textbf {\bibinfo
  {volume} {51}},\ \bibinfo {pages} {51} (\bibinfo {year} {1983})}\BibitemShut
  {NoStop}%
\bibitem [{\citenamefont {St{\"o}rmer}\ \emph {et~al.}(1986)\citenamefont
  {St{\"o}rmer}, \citenamefont {Eisenstein}, \citenamefont {Gossard},
  \citenamefont {Wiegmann},\ and\ \citenamefont
  {Baldwin}}]{stormer1986quantization}%
  \BibitemOpen
  \bibfield  {author} {\bibinfo {author} {\bibfnamefont {H.}~\bibnamefont
  {St{\"o}rmer}}, \bibinfo {author} {\bibfnamefont {J.}~\bibnamefont
  {Eisenstein}}, \bibinfo {author} {\bibfnamefont {A.}~\bibnamefont {Gossard}},
  \bibinfo {author} {\bibfnamefont {W.}~\bibnamefont {Wiegmann}},\ and\
  \bibinfo {author} {\bibfnamefont {K.}~\bibnamefont {Baldwin}},\ }\bibfield
  {title} {\bibinfo {title} {{Quantization of the Hall effect in an anisotropic
  three-dimensional electronic system}},\ }\href
  {https://doi.org/10.1103/PhysRevLett.56.85} {\bibfield  {journal} {\bibinfo
  {journal} {Phys. Rev. Lett.}\ }\textbf {\bibinfo {volume} {56}},\ \bibinfo
  {pages} {85} (\bibinfo {year} {1986})}\BibitemShut {NoStop}%
\bibitem [{\citenamefont {Gooth}\ \emph {et~al.}(2023)\citenamefont {Gooth},
  \citenamefont {Galeski},\ and\ \citenamefont {Meng}}]{gooth2023quantum}%
  \BibitemOpen
  \bibfield  {author} {\bibinfo {author} {\bibfnamefont {J.}~\bibnamefont
  {Gooth}}, \bibinfo {author} {\bibfnamefont {S.}~\bibnamefont {Galeski}},\
  and\ \bibinfo {author} {\bibfnamefont {T.}~\bibnamefont {Meng}},\ }\bibfield
  {title} {\bibinfo {title} {{Quantum-Hall physics and three dimensions}},\
  }\href {https://doi.org/10.1088/1361-6633/acb8c9} {\bibfield  {journal}
  {\bibinfo  {journal} {Rep. Prog. Phys.}\ }\textbf {\bibinfo {volume} {86}},\
  \bibinfo {pages} {044501} (\bibinfo {year} {2023})}\BibitemShut {NoStop}%
\bibitem [{\citenamefont {Hannahs}\ \emph {et~al.}(1989)\citenamefont
  {Hannahs}, \citenamefont {Brooks}, \citenamefont {Kang}, \citenamefont
  {Chiang},\ and\ \citenamefont {Chaikin}}]{hannahs1989quantum}%
  \BibitemOpen
  \bibfield  {author} {\bibinfo {author} {\bibfnamefont {S.}~\bibnamefont
  {Hannahs}}, \bibinfo {author} {\bibfnamefont {J.}~\bibnamefont {Brooks}},
  \bibinfo {author} {\bibfnamefont {W.}~\bibnamefont {Kang}}, \bibinfo {author}
  {\bibfnamefont {L.}~\bibnamefont {Chiang}},\ and\ \bibinfo {author}
  {\bibfnamefont {P.}~\bibnamefont {Chaikin}},\ }\bibfield  {title} {\bibinfo
  {title} {Quantum hall effect in a bulk crystal},\ }\href
  {https://doi.org/10.1103/PhysRevLett.63.1988} {\bibfield  {journal} {\bibinfo
   {journal} {Phys. Rev. Lett.}\ }\textbf {\bibinfo {volume} {63}},\ \bibinfo
  {pages} {1988} (\bibinfo {year} {1989})}\BibitemShut {NoStop}%
\bibitem [{\citenamefont {Cooper}\ \emph {et~al.}(1989)\citenamefont {Cooper},
  \citenamefont {Kang}, \citenamefont {Auban}, \citenamefont {Montambaux},
  \citenamefont {J{\'e}rome},\ and\ \citenamefont
  {Bechgaard}}]{cooper1989quantized}%
  \BibitemOpen
  \bibfield  {author} {\bibinfo {author} {\bibfnamefont {J.}~\bibnamefont
  {Cooper}}, \bibinfo {author} {\bibfnamefont {W.}~\bibnamefont {Kang}},
  \bibinfo {author} {\bibfnamefont {P.}~\bibnamefont {Auban}}, \bibinfo
  {author} {\bibfnamefont {G.}~\bibnamefont {Montambaux}}, \bibinfo {author}
  {\bibfnamefont {D.}~\bibnamefont {J{\'e}rome}},\ and\ \bibinfo {author}
  {\bibfnamefont {K.}~\bibnamefont {Bechgaard}},\ }\bibfield  {title} {\bibinfo
  {title} {{Quantized Hall effect and a new field-induced phase transition in
  the organic superconductor (TMTSF)$_2$PF$_6$}},\ }\href
  {https://doi.org/10.1103/PhysRevLett.63.1984} {\bibfield  {journal} {\bibinfo
   {journal} {Phys. Rev. Lett.}\ }\textbf {\bibinfo {volume} {63}},\ \bibinfo
  {pages} {1984} (\bibinfo {year} {1989})}\BibitemShut {NoStop}%
\bibitem [{\citenamefont {Hill}\ \emph {et~al.}(1998)\citenamefont {Hill},
  \citenamefont {Uji}, \citenamefont {Takashita}, \citenamefont {Terakura},
  \citenamefont {Terashima}, \citenamefont {Aoki}, \citenamefont {Brooks},
  \citenamefont {Fisk},\ and\ \citenamefont {Sarrao}}]{hill1998bulk}%
  \BibitemOpen
  \bibfield  {author} {\bibinfo {author} {\bibfnamefont {S.}~\bibnamefont
  {Hill}}, \bibinfo {author} {\bibfnamefont {S.}~\bibnamefont {Uji}}, \bibinfo
  {author} {\bibfnamefont {M.}~\bibnamefont {Takashita}}, \bibinfo {author}
  {\bibfnamefont {C.}~\bibnamefont {Terakura}}, \bibinfo {author}
  {\bibfnamefont {T.}~\bibnamefont {Terashima}}, \bibinfo {author}
  {\bibfnamefont {H.}~\bibnamefont {Aoki}}, \bibinfo {author} {\bibfnamefont
  {J.}~\bibnamefont {Brooks}}, \bibinfo {author} {\bibfnamefont
  {Z.}~\bibnamefont {Fisk}},\ and\ \bibinfo {author} {\bibfnamefont
  {J.}~\bibnamefont {Sarrao}},\ }\bibfield  {title} {\bibinfo {title} {{Bulk
  quantum Hall effect in $\eta$-Mo$_4$O$_{11}$}},\ }\href
  {https://doi.org/10.1103/PhysRevB.58.10778} {\bibfield  {journal} {\bibinfo
  {journal} {Phys. Rev. B}\ }\textbf {\bibinfo {volume} {58}},\ \bibinfo
  {pages} {10778} (\bibinfo {year} {1998})}\BibitemShut {NoStop}%
\bibitem [{\citenamefont {Kopelevich}\ \emph {et~al.}(2003)\citenamefont
  {Kopelevich}, \citenamefont {Torres}, \citenamefont {da~Silva}, \citenamefont
  {Mrowka}, \citenamefont {Kempa},\ and\ \citenamefont
  {Esquinazi}}]{kopelevich2003reentrant}%
  \BibitemOpen
  \bibfield  {author} {\bibinfo {author} {\bibfnamefont {Y.}~\bibnamefont
  {Kopelevich}}, \bibinfo {author} {\bibfnamefont {J.}~\bibnamefont {Torres}},
  \bibinfo {author} {\bibfnamefont {R.~R.}\ \bibnamefont {da~Silva}}, \bibinfo
  {author} {\bibfnamefont {F.}~\bibnamefont {Mrowka}}, \bibinfo {author}
  {\bibfnamefont {H.}~\bibnamefont {Kempa}},\ and\ \bibinfo {author}
  {\bibfnamefont {P.}~\bibnamefont {Esquinazi}},\ }\bibfield  {title} {\bibinfo
  {title} {Reentrant metallic behavior of graphite in the quantum limit},\
  }\href {https://doi.org/10.1103/PhysRevLett.90.156402} {\bibfield  {journal}
  {\bibinfo  {journal} {Phys. Rev. Lett.}\ }\textbf {\bibinfo {volume} {90}},\
  \bibinfo {pages} {156402} (\bibinfo {year} {2003})}\BibitemShut {NoStop}%
\bibitem [{\citenamefont {Kopelevich}\ \emph {et~al.}(2009)\citenamefont
  {Kopelevich}, \citenamefont {Raquet}, \citenamefont {Goiran}, \citenamefont
  {Escoffier}, \citenamefont {Da~Silva}, \citenamefont {Pantoja}, \citenamefont
  {Luk’yanchuk}, \citenamefont {Sinchenko},\ and\ \citenamefont
  {Monceau}}]{kopelevich2009searching}%
  \BibitemOpen
  \bibfield  {author} {\bibinfo {author} {\bibfnamefont {Y.}~\bibnamefont
  {Kopelevich}}, \bibinfo {author} {\bibfnamefont {B.}~\bibnamefont {Raquet}},
  \bibinfo {author} {\bibfnamefont {M.}~\bibnamefont {Goiran}}, \bibinfo
  {author} {\bibfnamefont {W.}~\bibnamefont {Escoffier}}, \bibinfo {author}
  {\bibfnamefont {R.}~\bibnamefont {Da~Silva}}, \bibinfo {author}
  {\bibfnamefont {J.~M.}\ \bibnamefont {Pantoja}}, \bibinfo {author}
  {\bibfnamefont {I.~A.}\ \bibnamefont {Luk’yanchuk}}, \bibinfo {author}
  {\bibfnamefont {A.}~\bibnamefont {Sinchenko}},\ and\ \bibinfo {author}
  {\bibfnamefont {P.}~\bibnamefont {Monceau}},\ }\bibfield  {title} {\bibinfo
  {title} {Searching for the fractional quantum hall effect in graphite},\
  }\href {https://doi.org/10.1103/PhysRevLett.103.116802} {\bibfield  {journal}
  {\bibinfo  {journal} {Phys. Rev. Lett.}\ }\textbf {\bibinfo {volume} {103}},\
  \bibinfo {pages} {116802} (\bibinfo {year} {2009})}\BibitemShut {NoStop}%
\bibitem [{\citenamefont {Yaguchi}\ and\ \citenamefont
  {Singleton}(2009)}]{yaguchi2009high}%
  \BibitemOpen
  \bibfield  {author} {\bibinfo {author} {\bibfnamefont {H.}~\bibnamefont
  {Yaguchi}}\ and\ \bibinfo {author} {\bibfnamefont {J.}~\bibnamefont
  {Singleton}},\ }\bibfield  {title} {\bibinfo {title} {A
  high-magnetic-field-induced density-wave state in graphite},\ }\href
  {https://doi.org/10.1088/0953-8984/21/34/344207} {\bibfield  {journal}
  {\bibinfo  {journal} {J. Condens. Matter Phys.}\ }\textbf {\bibinfo {volume}
  {21}},\ \bibinfo {pages} {344207} (\bibinfo {year} {2009})}\BibitemShut
  {NoStop}%
\bibitem [{\citenamefont {Cao}\ \emph {et~al.}(2012)\citenamefont {Cao},
  \citenamefont {Tian}, \citenamefont {Miotkowski}, \citenamefont {Shen},
  \citenamefont {Hu}, \citenamefont {Qiao},\ and\ \citenamefont
  {Chen}}]{cao2012quantized}%
  \BibitemOpen
  \bibfield  {author} {\bibinfo {author} {\bibfnamefont {H.}~\bibnamefont
  {Cao}}, \bibinfo {author} {\bibfnamefont {J.}~\bibnamefont {Tian}}, \bibinfo
  {author} {\bibfnamefont {I.}~\bibnamefont {Miotkowski}}, \bibinfo {author}
  {\bibfnamefont {T.}~\bibnamefont {Shen}}, \bibinfo {author} {\bibfnamefont
  {J.}~\bibnamefont {Hu}}, \bibinfo {author} {\bibfnamefont {S.}~\bibnamefont
  {Qiao}},\ and\ \bibinfo {author} {\bibfnamefont {Y.~P.}\ \bibnamefont
  {Chen}},\ }\bibfield  {title} {\bibinfo {title} {{Quantized Hall effect and
  Shubnikov--de Haas oscillations in highly doped Bi$_2$Se$_3$: Evidence for
  layered transport of bulk carriers}},\ }\href
  {https://doi.org/10.1103/PhysRevLett.108.216803} {\bibfield  {journal}
  {\bibinfo  {journal} {Phys. Rev. Lett.}\ }\textbf {\bibinfo {volume} {108}},\
  \bibinfo {pages} {216803} (\bibinfo {year} {2012})}\BibitemShut {NoStop}%
\bibitem [{\citenamefont {Masuda}\ \emph {et~al.}(2016)\citenamefont {Masuda},
  \citenamefont {Sakai}, \citenamefont {Tokunaga}, \citenamefont {Yamasaki},
  \citenamefont {Miyake}, \citenamefont {Shiogai}, \citenamefont {Nakamura},
  \citenamefont {Awaji}, \citenamefont {Tsukazaki}, \citenamefont {Nakao} \emph
  {et~al.}}]{masuda2016quantum}%
  \BibitemOpen
  \bibfield  {author} {\bibinfo {author} {\bibfnamefont {H.}~\bibnamefont
  {Masuda}}, \bibinfo {author} {\bibfnamefont {H.}~\bibnamefont {Sakai}},
  \bibinfo {author} {\bibfnamefont {M.}~\bibnamefont {Tokunaga}}, \bibinfo
  {author} {\bibfnamefont {Y.}~\bibnamefont {Yamasaki}}, \bibinfo {author}
  {\bibfnamefont {A.}~\bibnamefont {Miyake}}, \bibinfo {author} {\bibfnamefont
  {J.}~\bibnamefont {Shiogai}}, \bibinfo {author} {\bibfnamefont
  {S.}~\bibnamefont {Nakamura}}, \bibinfo {author} {\bibfnamefont
  {S.}~\bibnamefont {Awaji}}, \bibinfo {author} {\bibfnamefont
  {A.}~\bibnamefont {Tsukazaki}}, \bibinfo {author} {\bibfnamefont
  {H.}~\bibnamefont {Nakao}}, \emph {et~al.},\ }\bibfield  {title} {\bibinfo
  {title} {{Quantum Hall effect in a bulk antiferromagnet EuMnBi$_2$ with
  magnetically confined two-dimensional Dirac fermions}},\ }\href
  {https://doi.org/10.1126/sciadv.1501117} {\bibfield  {journal} {\bibinfo
  {journal} {Sci. Adv.}\ }\textbf {\bibinfo {volume} {2}},\ \bibinfo {pages}
  {e1501117} (\bibinfo {year} {2016})}\BibitemShut {NoStop}%
\bibitem [{\citenamefont {Sakai}\ \emph {et~al.}(2020)\citenamefont {Sakai},
  \citenamefont {Fujimura}, \citenamefont {Sakuragi}, \citenamefont {Ochi},
  \citenamefont {Kurihara}, \citenamefont {Miyake}, \citenamefont {Tokunaga},
  \citenamefont {Kojima}, \citenamefont {Hashizume}, \citenamefont {Muro} \emph
  {et~al.}}]{sakai2020bulk}%
  \BibitemOpen
  \bibfield  {author} {\bibinfo {author} {\bibfnamefont {H.}~\bibnamefont
  {Sakai}}, \bibinfo {author} {\bibfnamefont {H.}~\bibnamefont {Fujimura}},
  \bibinfo {author} {\bibfnamefont {S.}~\bibnamefont {Sakuragi}}, \bibinfo
  {author} {\bibfnamefont {M.}~\bibnamefont {Ochi}}, \bibinfo {author}
  {\bibfnamefont {R.}~\bibnamefont {Kurihara}}, \bibinfo {author}
  {\bibfnamefont {A.}~\bibnamefont {Miyake}}, \bibinfo {author} {\bibfnamefont
  {M.}~\bibnamefont {Tokunaga}}, \bibinfo {author} {\bibfnamefont
  {T.}~\bibnamefont {Kojima}}, \bibinfo {author} {\bibfnamefont
  {D.}~\bibnamefont {Hashizume}}, \bibinfo {author} {\bibfnamefont
  {T.}~\bibnamefont {Muro}}, \emph {et~al.},\ }\bibfield  {title} {\bibinfo
  {title} {{Bulk quantum Hall effect of spin-valley coupled Dirac fermions in
  the polar antiferromagnet BaMnSb$_2$}},\ }\href
  {https://doi.org/10.1103/PhysRevB.101.081104} {\bibfield  {journal} {\bibinfo
   {journal} {Phys. Rev. B}\ }\textbf {\bibinfo {volume} {101}},\ \bibinfo
  {pages} {081104} (\bibinfo {year} {2020})}\BibitemShut {NoStop}%
\bibitem [{\citenamefont {Liu}\ \emph {et~al.}(2017)\citenamefont {Liu},
  \citenamefont {Long}, \citenamefont {Zhao}, \citenamefont {Nie},
  \citenamefont {Zhang}, \citenamefont {Weng}, \citenamefont {Jin},
  \citenamefont {Li}, \citenamefont {Liu}, \citenamefont {Long} \emph
  {et~al.}}]{liu2017superconductivity}%
  \BibitemOpen
  \bibfield  {author} {\bibinfo {author} {\bibfnamefont {Y.}~\bibnamefont
  {Liu}}, \bibinfo {author} {\bibfnamefont {Y.}~\bibnamefont {Long}}, \bibinfo
  {author} {\bibfnamefont {L.}~\bibnamefont {Zhao}}, \bibinfo {author}
  {\bibfnamefont {S.}~\bibnamefont {Nie}}, \bibinfo {author} {\bibfnamefont
  {S.}~\bibnamefont {Zhang}}, \bibinfo {author} {\bibfnamefont
  {Y.}~\bibnamefont {Weng}}, \bibinfo {author} {\bibfnamefont {M.}~\bibnamefont
  {Jin}}, \bibinfo {author} {\bibfnamefont {W.}~\bibnamefont {Li}}, \bibinfo
  {author} {\bibfnamefont {Q.}~\bibnamefont {Liu}}, \bibinfo {author}
  {\bibfnamefont {Y.}~\bibnamefont {Long}}, \emph {et~al.},\ }\bibfield
  {title} {\bibinfo {title} {{Superconductivity in HfTe$_5$ across weak to
  strong topological insulator transition induced via pressures}},\ }\href
  {https://doi.org/10.1038/srep44367} {\bibfield  {journal} {\bibinfo
  {journal} {Sci. Rep.}\ }\textbf {\bibinfo {volume} {7}},\ \bibinfo {pages}
  {44367} (\bibinfo {year} {2017})}\BibitemShut {NoStop}%
\bibitem [{\citenamefont {Tang}\ \emph {et~al.}(2019)\citenamefont {Tang},
  \citenamefont {Ren}, \citenamefont {Wang}, \citenamefont {Zhong},
  \citenamefont {Schneeloch}, \citenamefont {Yang}, \citenamefont {Yang},
  \citenamefont {Lee}, \citenamefont {Gu}, \citenamefont {Qiao} \emph
  {et~al.}}]{tang2019three}%
  \BibitemOpen
  \bibfield  {author} {\bibinfo {author} {\bibfnamefont {F.}~\bibnamefont
  {Tang}}, \bibinfo {author} {\bibfnamefont {Y.}~\bibnamefont {Ren}}, \bibinfo
  {author} {\bibfnamefont {P.}~\bibnamefont {Wang}}, \bibinfo {author}
  {\bibfnamefont {R.}~\bibnamefont {Zhong}}, \bibinfo {author} {\bibfnamefont
  {J.}~\bibnamefont {Schneeloch}}, \bibinfo {author} {\bibfnamefont {S.~A.}\
  \bibnamefont {Yang}}, \bibinfo {author} {\bibfnamefont {K.}~\bibnamefont
  {Yang}}, \bibinfo {author} {\bibfnamefont {P.~A.}\ \bibnamefont {Lee}},
  \bibinfo {author} {\bibfnamefont {G.}~\bibnamefont {Gu}}, \bibinfo {author}
  {\bibfnamefont {Z.}~\bibnamefont {Qiao}}, \emph {et~al.},\ }\bibfield
  {title} {\bibinfo {title} {{Three-dimensional quantum Hall effect and
  metal--insulator transition in ZrTe$_5$}},\ }\href
  {https://doi.org/10.1038/s41586-019-1180-9} {\bibfield  {journal} {\bibinfo
  {journal} {Nature}\ }\textbf {\bibinfo {volume} {569}},\ \bibinfo {pages}
  {537} (\bibinfo {year} {2019})}\BibitemShut {NoStop}%
\bibitem [{\citenamefont {Wang}\ \emph {et~al.}(2020)\citenamefont {Wang},
  \citenamefont {Ren}, \citenamefont {Tang}, \citenamefont {Wang},
  \citenamefont {Hou}, \citenamefont {Zeng}, \citenamefont {Zhang},\ and\
  \citenamefont {Qiao}}]{wang2020approaching}%
  \BibitemOpen
  \bibfield  {author} {\bibinfo {author} {\bibfnamefont {P.}~\bibnamefont
  {Wang}}, \bibinfo {author} {\bibfnamefont {Y.}~\bibnamefont {Ren}}, \bibinfo
  {author} {\bibfnamefont {F.}~\bibnamefont {Tang}}, \bibinfo {author}
  {\bibfnamefont {P.}~\bibnamefont {Wang}}, \bibinfo {author} {\bibfnamefont
  {T.}~\bibnamefont {Hou}}, \bibinfo {author} {\bibfnamefont {H.}~\bibnamefont
  {Zeng}}, \bibinfo {author} {\bibfnamefont {L.}~\bibnamefont {Zhang}},\ and\
  \bibinfo {author} {\bibfnamefont {Z.}~\bibnamefont {Qiao}},\ }\bibfield
  {title} {\bibinfo {title} {{Approaching three-dimensional quantum Hall effect
  in bulk HfTe$_{5}$}},\ }\href {https://doi.org/10.1103/PhysRevB.101.161201}
  {\bibfield  {journal} {\bibinfo  {journal} {Phys. Rev. B}\ }\textbf {\bibinfo
  {volume} {101}},\ \bibinfo {pages} {161201} (\bibinfo {year}
  {2020})}\BibitemShut {NoStop}%
\bibitem [{\citenamefont {Galeski}\ \emph {et~al.}(2020)\citenamefont
  {Galeski}, \citenamefont {Zhao}, \citenamefont {Wawrzy{\'n}czak},
  \citenamefont {Meng}, \citenamefont {F{\"o}rster}, \citenamefont {Lozano},
  \citenamefont {Honnali}, \citenamefont {Lamba}, \citenamefont {Ehmcke},
  \citenamefont {Markou} \emph {et~al.}}]{galeski2020unconventional}%
  \BibitemOpen
  \bibfield  {author} {\bibinfo {author} {\bibfnamefont {S.}~\bibnamefont
  {Galeski}}, \bibinfo {author} {\bibfnamefont {X.}~\bibnamefont {Zhao}},
  \bibinfo {author} {\bibfnamefont {R.}~\bibnamefont {Wawrzy{\'n}czak}},
  \bibinfo {author} {\bibfnamefont {T.}~\bibnamefont {Meng}}, \bibinfo {author}
  {\bibfnamefont {T.}~\bibnamefont {F{\"o}rster}}, \bibinfo {author}
  {\bibfnamefont {P.}~\bibnamefont {Lozano}}, \bibinfo {author} {\bibfnamefont
  {S.}~\bibnamefont {Honnali}}, \bibinfo {author} {\bibfnamefont
  {N.}~\bibnamefont {Lamba}}, \bibinfo {author} {\bibfnamefont
  {T.}~\bibnamefont {Ehmcke}}, \bibinfo {author} {\bibfnamefont
  {A.}~\bibnamefont {Markou}}, \emph {et~al.},\ }\bibfield  {title} {\bibinfo
  {title} {{Unconventional Hall response in the quantum limit of HfTe$_5$}},\
  }\href {https://doi.org/10.1038/s41467-020-19773-y} {\bibfield  {journal}
  {\bibinfo  {journal} {Nat. Comm.}\ }\textbf {\bibinfo {volume} {11}},\
  \bibinfo {pages} {5926} (\bibinfo {year} {2020})}\BibitemShut {NoStop}%
\bibitem [{\citenamefont {Galeski}\ \emph {et~al.}(2021)\citenamefont
  {Galeski}, \citenamefont {Ehmcke}, \citenamefont {Wawrzy{\'n}czak},
  \citenamefont {Lozano}, \citenamefont {Cho}, \citenamefont {Sharma},
  \citenamefont {Das}, \citenamefont {K{\"u}ster}, \citenamefont {Sessi},
  \citenamefont {Brando} \emph {et~al.}}]{galeski2021origin}%
  \BibitemOpen
  \bibfield  {author} {\bibinfo {author} {\bibfnamefont {S.}~\bibnamefont
  {Galeski}}, \bibinfo {author} {\bibfnamefont {T.}~\bibnamefont {Ehmcke}},
  \bibinfo {author} {\bibfnamefont {R.}~\bibnamefont {Wawrzy{\'n}czak}},
  \bibinfo {author} {\bibfnamefont {P.~M.}\ \bibnamefont {Lozano}}, \bibinfo
  {author} {\bibfnamefont {K.}~\bibnamefont {Cho}}, \bibinfo {author}
  {\bibfnamefont {A.}~\bibnamefont {Sharma}}, \bibinfo {author} {\bibfnamefont
  {S.}~\bibnamefont {Das}}, \bibinfo {author} {\bibfnamefont {F.}~\bibnamefont
  {K{\"u}ster}}, \bibinfo {author} {\bibfnamefont {P.}~\bibnamefont {Sessi}},
  \bibinfo {author} {\bibfnamefont {M.}~\bibnamefont {Brando}}, \emph
  {et~al.},\ }\bibfield  {title} {\bibinfo {title} {{Origin of the
  quasi-quantized Hall effect in ZrTe$_5$}},\ }\href
  {https://doi.org/10.1038/s41467-021-23435-y} {\bibfield  {journal} {\bibinfo
  {journal} {Nat. Comm.}\ }\textbf {\bibinfo {volume} {12}},\ \bibinfo {pages}
  {3197} (\bibinfo {year} {2021})}\BibitemShut {NoStop}%
\bibitem [{\citenamefont {Furuseth}\ \emph {et~al.}(1973)\citenamefont
  {Furuseth}, \citenamefont {Brattas},\ and\ \citenamefont
  {Kjekshus}}]{furuseth1973crystal}%
  \BibitemOpen
  \bibfield  {author} {\bibinfo {author} {\bibfnamefont {S.}~\bibnamefont
  {Furuseth}}, \bibinfo {author} {\bibfnamefont {L.}~\bibnamefont {Brattas}},\
  and\ \bibinfo {author} {\bibfnamefont {A.}~\bibnamefont {Kjekshus}},\
  }\bibfield  {title} {\bibinfo {title} {{Crystal structure of HfTe$_5$}},\
  }\href {https://doi.org/10.3891/acta.chem.scand.27-2367} {\bibfield
  {journal} {\bibinfo  {journal} {Acta Chem. Scand.}\ }\textbf {\bibinfo
  {volume} {27}},\ \bibinfo {pages} {2367} (\bibinfo {year}
  {1973})}\BibitemShut {NoStop}%
\bibitem [{\citenamefont {Fjellv{\aa}g}\ and\ \citenamefont
  {Kjekshus}(1986)}]{fjellvaag1986structural}%
  \BibitemOpen
  \bibfield  {author} {\bibinfo {author} {\bibfnamefont {H.}~\bibnamefont
  {Fjellv{\aa}g}}\ and\ \bibinfo {author} {\bibfnamefont {A.}~\bibnamefont
  {Kjekshus}},\ }\bibfield  {title} {\bibinfo {title} {{Structural properties
  of ZrTe$_5$ and HfTe$_5$ as seen by powder diffraction}},\ }\href
  {https://doi.org/10.1016/0038-1098(86)90536-3} {\bibfield  {journal}
  {\bibinfo  {journal} {Solid State Commun.}\ }\textbf {\bibinfo {volume}
  {60}},\ \bibinfo {pages} {91} (\bibinfo {year} {1986})}\BibitemShut {NoStop}%
\bibitem [{\citenamefont {Fan}\ \emph {et~al.}(2017)\citenamefont {Fan},
  \citenamefont {Liang}, \citenamefont {Chen}, \citenamefont {Yao},\ and\
  \citenamefont {Zhou}}]{fan2017transition}%
  \BibitemOpen
  \bibfield  {author} {\bibinfo {author} {\bibfnamefont {Z.}~\bibnamefont
  {Fan}}, \bibinfo {author} {\bibfnamefont {Q.-F.}\ \bibnamefont {Liang}},
  \bibinfo {author} {\bibfnamefont {Y.}~\bibnamefont {Chen}}, \bibinfo {author}
  {\bibfnamefont {S.-H.}\ \bibnamefont {Yao}},\ and\ \bibinfo {author}
  {\bibfnamefont {J.}~\bibnamefont {Zhou}},\ }\bibfield  {title} {\bibinfo
  {title} {{Transition between strong and weak topological insulator in
  ZrTe$_{5}$ and HfTe$_{5}$}},\ }\href {https://doi.org/10.1038/srep45667}
  {\bibfield  {journal} {\bibinfo  {journal} {Sci. Rep.}\ }\textbf {\bibinfo
  {volume} {7}},\ \bibinfo {pages} {45667} (\bibinfo {year}
  {2017})}\BibitemShut {NoStop}%
\bibitem [{\citenamefont {Facio}\ \emph {et~al.}(2023)\citenamefont {Facio},
  \citenamefont {Nocerino}, \citenamefont {Fulga}, \citenamefont {Wawrzynczak},
  \citenamefont {Brown}, \citenamefont {Gu}, \citenamefont {Li}, \citenamefont
  {Mansson}, \citenamefont {Sassa}, \citenamefont {Ivashko} \emph
  {et~al.}}]{facio2023engineering}%
  \BibitemOpen
  \bibfield  {author} {\bibinfo {author} {\bibfnamefont {J.~I.}\ \bibnamefont
  {Facio}}, \bibinfo {author} {\bibfnamefont {E.}~\bibnamefont {Nocerino}},
  \bibinfo {author} {\bibfnamefont {I.~C.}\ \bibnamefont {Fulga}}, \bibinfo
  {author} {\bibfnamefont {R.}~\bibnamefont {Wawrzynczak}}, \bibinfo {author}
  {\bibfnamefont {J.}~\bibnamefont {Brown}}, \bibinfo {author} {\bibfnamefont
  {G.}~\bibnamefont {Gu}}, \bibinfo {author} {\bibfnamefont {Q.}~\bibnamefont
  {Li}}, \bibinfo {author} {\bibfnamefont {M.}~\bibnamefont {Mansson}},
  \bibinfo {author} {\bibfnamefont {Y.}~\bibnamefont {Sassa}}, \bibinfo
  {author} {\bibfnamefont {O.}~\bibnamefont {Ivashko}}, \emph {et~al.},\
  }\bibfield  {title} {\bibinfo {title} {{Engineering a pure Dirac regime in
  ZrTe$_5$}},\ }\href {https://doi.org/10.21468/SciPostPhys.14.4.066}
  {\bibfield  {journal} {\bibinfo  {journal} {SciPost Phys.}\ }\textbf
  {\bibinfo {volume} {14}},\ \bibinfo {pages} {066} (\bibinfo {year}
  {2023})}\BibitemShut {NoStop}%
\bibitem [{\citenamefont {Lifshitz}(1960)}]{lifshitz1960anomalies}%
  \BibitemOpen
  \bibfield  {author} {\bibinfo {author} {\bibfnamefont {I.}~\bibnamefont
  {Lifshitz}},\ }\bibfield  {title} {\bibinfo {title} {Anomalies of electron
  characteristics of a metal in the high pressure region},\ }\href@noop {}
  {\bibfield  {journal} {\bibinfo  {journal} {Sov. Phys. JETP}\ }\textbf
  {\bibinfo {volume} {11}},\ \bibinfo {pages} {1130} (\bibinfo {year}
  {1960})}\BibitemShut {NoStop}%
\bibitem [{\citenamefont {Zhang}\ \emph
  {et~al.}(2017{\natexlab{a}})\citenamefont {Zhang}, \citenamefont {Wang},
  \citenamefont {Yu}, \citenamefont {Liu}, \citenamefont {Liang}, \citenamefont
  {Huang}, \citenamefont {Nie}, \citenamefont {Sun}, \citenamefont {Zhang},
  \citenamefont {Shen} \emph {et~al.}}]{zhang2017electronic}%
  \BibitemOpen
  \bibfield  {author} {\bibinfo {author} {\bibfnamefont {Y.}~\bibnamefont
  {Zhang}}, \bibinfo {author} {\bibfnamefont {C.}~\bibnamefont {Wang}},
  \bibinfo {author} {\bibfnamefont {L.}~\bibnamefont {Yu}}, \bibinfo {author}
  {\bibfnamefont {G.}~\bibnamefont {Liu}}, \bibinfo {author} {\bibfnamefont
  {A.}~\bibnamefont {Liang}}, \bibinfo {author} {\bibfnamefont
  {J.}~\bibnamefont {Huang}}, \bibinfo {author} {\bibfnamefont
  {S.}~\bibnamefont {Nie}}, \bibinfo {author} {\bibfnamefont {X.}~\bibnamefont
  {Sun}}, \bibinfo {author} {\bibfnamefont {Y.}~\bibnamefont {Zhang}}, \bibinfo
  {author} {\bibfnamefont {B.}~\bibnamefont {Shen}}, \emph {et~al.},\
  }\bibfield  {title} {\bibinfo {title} {{Electronic evidence of
  temperature-induced Lifshitz transition and topological nature in
  ZrTe$_5$}},\ }\href {https://doi.org/10.1038/ncomms15512} {\bibfield
  {journal} {\bibinfo  {journal} {Nat. Comm.}\ }\textbf {\bibinfo {volume}
  {8}},\ \bibinfo {pages} {15512} (\bibinfo {year}
  {2017}{\natexlab{a}})}\BibitemShut {NoStop}%
\bibitem [{\citenamefont {Zhang}\ \emph
  {et~al.}(2017{\natexlab{b}})\citenamefont {Zhang}, \citenamefont {Wang},
  \citenamefont {Liu}, \citenamefont {Liang}, \citenamefont {Zhao},
  \citenamefont {Huang}, \citenamefont {Gao}, \citenamefont {Shen},
  \citenamefont {Liu}, \citenamefont {Hu} \emph
  {et~al.}}]{zhang2017temperature}%
  \BibitemOpen
  \bibfield  {author} {\bibinfo {author} {\bibfnamefont {Y.}~\bibnamefont
  {Zhang}}, \bibinfo {author} {\bibfnamefont {C.}~\bibnamefont {Wang}},
  \bibinfo {author} {\bibfnamefont {G.}~\bibnamefont {Liu}}, \bibinfo {author}
  {\bibfnamefont {A.}~\bibnamefont {Liang}}, \bibinfo {author} {\bibfnamefont
  {L.}~\bibnamefont {Zhao}}, \bibinfo {author} {\bibfnamefont {J.}~\bibnamefont
  {Huang}}, \bibinfo {author} {\bibfnamefont {Q.}~\bibnamefont {Gao}}, \bibinfo
  {author} {\bibfnamefont {B.}~\bibnamefont {Shen}}, \bibinfo {author}
  {\bibfnamefont {J.}~\bibnamefont {Liu}}, \bibinfo {author} {\bibfnamefont
  {C.}~\bibnamefont {Hu}}, \emph {et~al.},\ }\bibfield  {title} {\bibinfo
  {title} {{Temperature-induced Lifshitz transition in topological insulator
  candidate HfTe$_5$}},\ }\href {https://doi.org/10.1038/ncomms15512}
  {\bibfield  {journal} {\bibinfo  {journal} {Sci. Bull.}\ }\textbf {\bibinfo
  {volume} {62}},\ \bibinfo {pages} {950} (\bibinfo {year}
  {2017}{\natexlab{b}})}\BibitemShut {NoStop}%
\bibitem [{\citenamefont {Qin}\ \emph {et~al.}(2020)\citenamefont {Qin},
  \citenamefont {Li}, \citenamefont {Du}, \citenamefont {Wang}, \citenamefont
  {Zhang}, \citenamefont {Yu}, \citenamefont {Lu}, \citenamefont {Xie} \emph
  {et~al.}}]{qin2020theory}%
  \BibitemOpen
  \bibfield  {author} {\bibinfo {author} {\bibfnamefont {F.}~\bibnamefont
  {Qin}}, \bibinfo {author} {\bibfnamefont {S.}~\bibnamefont {Li}}, \bibinfo
  {author} {\bibfnamefont {Z.}~\bibnamefont {Du}}, \bibinfo {author}
  {\bibfnamefont {C.}~\bibnamefont {Wang}}, \bibinfo {author} {\bibfnamefont
  {W.}~\bibnamefont {Zhang}}, \bibinfo {author} {\bibfnamefont
  {D.}~\bibnamefont {Yu}}, \bibinfo {author} {\bibfnamefont {H.-Z.}\
  \bibnamefont {Lu}}, \bibinfo {author} {\bibfnamefont {X.}~\bibnamefont
  {Xie}}, \emph {et~al.},\ }\bibfield  {title} {\bibinfo {title} {{Theory for
  the charge-density-wave mechanism of 3D quantum Hall effect}},\ }\href
  {https://doi.org/10.1103/PhysRevLett.125.206601} {\bibfield  {journal}
  {\bibinfo  {journal} {Phys. Rev. Lett.}\ }\textbf {\bibinfo {volume} {125}},\
  \bibinfo {pages} {206601} (\bibinfo {year} {2020})}\BibitemShut {NoStop}%
\bibitem [{\citenamefont {Qi}\ \emph {et~al.}(2016)\citenamefont {Qi},
  \citenamefont {Shi}, \citenamefont {Naumov}, \citenamefont {Kumar},
  \citenamefont {Schnelle}, \citenamefont {Barkalov}, \citenamefont {Shekhar},
  \citenamefont {Borrmann}, \citenamefont {Felser}, \citenamefont {Yan} \emph
  {et~al.}}]{qi2016pressure}%
  \BibitemOpen
  \bibfield  {author} {\bibinfo {author} {\bibfnamefont {Y.}~\bibnamefont
  {Qi}}, \bibinfo {author} {\bibfnamefont {W.}~\bibnamefont {Shi}}, \bibinfo
  {author} {\bibfnamefont {P.~G.}\ \bibnamefont {Naumov}}, \bibinfo {author}
  {\bibfnamefont {N.}~\bibnamefont {Kumar}}, \bibinfo {author} {\bibfnamefont
  {W.}~\bibnamefont {Schnelle}}, \bibinfo {author} {\bibfnamefont
  {O.}~\bibnamefont {Barkalov}}, \bibinfo {author} {\bibfnamefont
  {C.}~\bibnamefont {Shekhar}}, \bibinfo {author} {\bibfnamefont
  {H.}~\bibnamefont {Borrmann}}, \bibinfo {author} {\bibfnamefont
  {C.}~\bibnamefont {Felser}}, \bibinfo {author} {\bibfnamefont
  {B.}~\bibnamefont {Yan}}, \emph {et~al.},\ }\bibfield  {title} {\bibinfo
  {title} {{Pressure-driven superconductivity in the transition-metal
  pentatelluride HfTe$_{5}$}},\ }\href
  {https://doi.org/10.1103/PhysRevB.94.054517} {\bibfield  {journal} {\bibinfo
  {journal} {Phys. Rev. B}\ }\textbf {\bibinfo {volume} {94}},\ \bibinfo
  {pages} {054517} (\bibinfo {year} {2016})}\BibitemShut {NoStop}%
\bibitem [{\citenamefont {Liu}\ \emph {et~al.}(2024)\citenamefont {Liu},
  \citenamefont {Zhou}, \citenamefont {Yepez~Rodriguez}, \citenamefont
  {Delmont}, \citenamefont {Welser}, \citenamefont {Ho}, \citenamefont
  {Sirica}, \citenamefont {McClure}, \citenamefont {Vilmercati}, \citenamefont
  {Ziller} \emph {et~al.}}]{liu2023controllable}%
  \BibitemOpen
  \bibfield  {author} {\bibinfo {author} {\bibfnamefont {J.}~\bibnamefont
  {Liu}}, \bibinfo {author} {\bibfnamefont {Y.}~\bibnamefont {Zhou}}, \bibinfo
  {author} {\bibfnamefont {S.}~\bibnamefont {Yepez~Rodriguez}}, \bibinfo
  {author} {\bibfnamefont {M.~A.}\ \bibnamefont {Delmont}}, \bibinfo {author}
  {\bibfnamefont {R.~A.}\ \bibnamefont {Welser}}, \bibinfo {author}
  {\bibfnamefont {T.}~\bibnamefont {Ho}}, \bibinfo {author} {\bibfnamefont
  {N.}~\bibnamefont {Sirica}}, \bibinfo {author} {\bibfnamefont
  {K.}~\bibnamefont {McClure}}, \bibinfo {author} {\bibfnamefont
  {P.}~\bibnamefont {Vilmercati}}, \bibinfo {author} {\bibfnamefont {J.~W.}\
  \bibnamefont {Ziller}}, \emph {et~al.},\ }\bibfield  {title} {\bibinfo
  {title} {{Controllable strain-driven topological phase transition and
  dominant surface-state transport in HfTe$_{5}$}},\ }\href
  {https://doi.org/10.1038/s41467-023-44547-7} {\bibfield  {journal} {\bibinfo
  {journal} {Nature Communications}\ }\textbf {\bibinfo {volume} {15}},\
  \bibinfo {pages} {332} (\bibinfo {year} {2024})}\BibitemShut {NoStop}%
\bibitem [{\citenamefont {Jo}\ \emph {et~al.}(2023)\citenamefont {Jo},
  \citenamefont {Ashour}, \citenamefont {Shu}, \citenamefont {Jozwiak},
  \citenamefont {Bostwick}, \citenamefont {Ryu}, \citenamefont {Sun},
  \citenamefont {Kong}, \citenamefont {Griffin},\ and\ \citenamefont
  {Rotenberg}}]{jo2023effects}%
  \BibitemOpen
  \bibfield  {author} {\bibinfo {author} {\bibfnamefont {N.~H.}\ \bibnamefont
  {Jo}}, \bibinfo {author} {\bibfnamefont {O.~A.}\ \bibnamefont {Ashour}},
  \bibinfo {author} {\bibfnamefont {Z.}~\bibnamefont {Shu}}, \bibinfo {author}
  {\bibfnamefont {C.}~\bibnamefont {Jozwiak}}, \bibinfo {author} {\bibfnamefont
  {A.}~\bibnamefont {Bostwick}}, \bibinfo {author} {\bibfnamefont {S.~H.}\
  \bibnamefont {Ryu}}, \bibinfo {author} {\bibfnamefont {K.}~\bibnamefont
  {Sun}}, \bibinfo {author} {\bibfnamefont {T.}~\bibnamefont {Kong}}, \bibinfo
  {author} {\bibfnamefont {S.~M.}\ \bibnamefont {Griffin}},\ and\ \bibinfo
  {author} {\bibfnamefont {E.}~\bibnamefont {Rotenberg}},\ }\bibfield  {title}
  {\bibinfo {title} {{On the effects of strain, defects, and interactions on
  the topological properties of HfTe$_{5}$}},\ }\bibfield  {journal} {\bibinfo
  {journal} {arXiv preprint arXiv:2303.10836}\ }\href
  {https://doi.org/10.48550/arXiv.2303.10836} {10.48550/arXiv.2303.10836}
  (\bibinfo {year} {2023})\BibitemShut {NoStop}%
\bibitem [{\citenamefont {Nicklas}(2015)}]{nicklas2015pressure}%
  \BibitemOpen
  \bibfield  {author} {\bibinfo {author} {\bibfnamefont {M.}~\bibnamefont
  {Nicklas}},\ }\bibfield  {title} {\bibinfo {title} {Pressure probes},\ }in\
  \href {https://doi.org/10.1007/978-3-662-44133-6_6} {\emph {\bibinfo
  {booktitle} {Strongly Correlated Systems}}}\ (\bibinfo  {publisher}
  {Springer},\ \bibinfo {year} {2015})\ pp.\ \bibinfo {pages}
  {173--204}\BibitemShut {NoStop}%
\bibitem [{\citenamefont {Fuller}\ \emph {et~al.}(1983)\citenamefont {Fuller},
  \citenamefont {Wolf}, \citenamefont {Wieting}, \citenamefont {LaCoe},
  \citenamefont {Chaikin},\ and\ \citenamefont {Huang}}]{fuller1983pressure}%
  \BibitemOpen
  \bibfield  {author} {\bibinfo {author} {\bibfnamefont {W.}~\bibnamefont
  {Fuller}}, \bibinfo {author} {\bibfnamefont {S.}~\bibnamefont {Wolf}},
  \bibinfo {author} {\bibfnamefont {T.}~\bibnamefont {Wieting}}, \bibinfo
  {author} {\bibfnamefont {R.}~\bibnamefont {LaCoe}}, \bibinfo {author}
  {\bibfnamefont {P.}~\bibnamefont {Chaikin}},\ and\ \bibinfo {author}
  {\bibfnamefont {C.}~\bibnamefont {Huang}},\ }\bibfield  {title} {\bibinfo
  {title} {{Pressure effects in HfTe$_{5}$ and ZrTe$_{5}$}},\ }\href
  {https://doi.org/10.1051/jphyscol/1983095} {\bibfield  {journal} {\bibinfo
  {journal} {J. Phys. Colloq.}\ }\textbf {\bibinfo {volume} {44}},\ \bibinfo
  {pages} {C3} (\bibinfo {year} {1983})}\BibitemShut {NoStop}%
\bibitem [{\citenamefont {Zhou}\ \emph {et~al.}(2016)\citenamefont {Zhou},
  \citenamefont {Wu}, \citenamefont {Ning}, \citenamefont {Li}, \citenamefont
  {Du}, \citenamefont {Chen}, \citenamefont {Zhang}, \citenamefont {Chi},
  \citenamefont {Wang}, \citenamefont {Zhu} \emph {et~al.}}]{zhou2016pressure}%
  \BibitemOpen
  \bibfield  {author} {\bibinfo {author} {\bibfnamefont {Y.}~\bibnamefont
  {Zhou}}, \bibinfo {author} {\bibfnamefont {J.}~\bibnamefont {Wu}}, \bibinfo
  {author} {\bibfnamefont {W.}~\bibnamefont {Ning}}, \bibinfo {author}
  {\bibfnamefont {N.}~\bibnamefont {Li}}, \bibinfo {author} {\bibfnamefont
  {Y.}~\bibnamefont {Du}}, \bibinfo {author} {\bibfnamefont {X.}~\bibnamefont
  {Chen}}, \bibinfo {author} {\bibfnamefont {R.}~\bibnamefont {Zhang}},
  \bibinfo {author} {\bibfnamefont {Z.}~\bibnamefont {Chi}}, \bibinfo {author}
  {\bibfnamefont {X.}~\bibnamefont {Wang}}, \bibinfo {author} {\bibfnamefont
  {X.}~\bibnamefont {Zhu}}, \emph {et~al.},\ }\bibfield  {title} {\bibinfo
  {title} {{Pressure-induced superconductivity in a three-dimensional
  topological material ZrTe$_5$}},\ }\href
  {https://doi.org/10.1073/pnas.1601262113} {\bibfield  {journal} {\bibinfo
  {journal} {Proc. Natl. Acad. Sci. U.S.A.}\ }\textbf {\bibinfo {volume}
  {113}},\ \bibinfo {pages} {2904} (\bibinfo {year} {2016})}\BibitemShut
  {NoStop}%
\bibitem [{\citenamefont {Seeger}(2013)}]{seeger2013semiconductor}%
  \BibitemOpen
  \bibfield  {author} {\bibinfo {author} {\bibfnamefont {K.}~\bibnamefont
  {Seeger}},\ }\href@noop {} {\emph {\bibinfo {title} {Semiconductor
  physics}}}\ (\bibinfo  {publisher} {Springer Science \& Business Media},\
  \bibinfo {year} {2013})\BibitemShut {NoStop}%
\bibitem [{\citenamefont {Weng}\ \emph {et~al.}(2014)\citenamefont {Weng},
  \citenamefont {Dai},\ and\ \citenamefont {Fang}}]{weng2014transition}%
  \BibitemOpen
  \bibfield  {author} {\bibinfo {author} {\bibfnamefont {H.}~\bibnamefont
  {Weng}}, \bibinfo {author} {\bibfnamefont {X.}~\bibnamefont {Dai}},\ and\
  \bibinfo {author} {\bibfnamefont {Z.}~\bibnamefont {Fang}},\ }\bibfield
  {title} {\bibinfo {title} {Transition-metal pentatelluride zrte 5 and hfte 5:
  A paradigm for large-gap quantum spin hall insulators},\ }\href
  {https://doi.org/10.1103/PhysRevX.4.011002} {\bibfield  {journal} {\bibinfo
  {journal} {Physical review X}\ }\textbf {\bibinfo {volume} {4}},\ \bibinfo
  {pages} {011002} (\bibinfo {year} {2014})}\BibitemShut {NoStop}%
\bibitem [{\citenamefont {Wu}\ \emph {et~al.}(2023)\citenamefont {Wu},
  \citenamefont {Shi}, \citenamefont {Du}, \citenamefont {Wang}, \citenamefont
  {Qin}, \citenamefont {Meng}, \citenamefont {Liu}, \citenamefont {Ma},
  \citenamefont {Yan}, \citenamefont {Ozerov} \emph
  {et~al.}}]{wu2023topological}%
  \BibitemOpen
  \bibfield  {author} {\bibinfo {author} {\bibfnamefont {W.}~\bibnamefont
  {Wu}}, \bibinfo {author} {\bibfnamefont {Z.}~\bibnamefont {Shi}}, \bibinfo
  {author} {\bibfnamefont {Y.}~\bibnamefont {Du}}, \bibinfo {author}
  {\bibfnamefont {Y.}~\bibnamefont {Wang}}, \bibinfo {author} {\bibfnamefont
  {F.}~\bibnamefont {Qin}}, \bibinfo {author} {\bibfnamefont {X.}~\bibnamefont
  {Meng}}, \bibinfo {author} {\bibfnamefont {B.}~\bibnamefont {Liu}}, \bibinfo
  {author} {\bibfnamefont {Y.}~\bibnamefont {Ma}}, \bibinfo {author}
  {\bibfnamefont {Z.}~\bibnamefont {Yan}}, \bibinfo {author} {\bibfnamefont
  {M.}~\bibnamefont {Ozerov}}, \emph {et~al.},\ }\bibfield  {title} {\bibinfo
  {title} {{Topological Lifshitz transition and one-dimensional Weyl mode in
  HfTe$_5$}},\ }\href {https://doi.org/10.1038/s41563-022-01364-5} {\bibfield
  {journal} {\bibinfo  {journal} {Nat. Mater.}\ }\textbf {\bibinfo {volume}
  {22}},\ \bibinfo {pages} {84} (\bibinfo {year} {2023})}\BibitemShut {NoStop}%
\bibitem [{SM()}]{SM}%
  \BibitemOpen
  \href@noop {} {\bibinfo  {journal} {The Supplemental Material at url provides
  additional information on X-ray diffraction measurements, the
  magnetoresistance and Hall resistivity. It also contains the longitudinal
  resistivity subtracted by a smooth background at 1.8~K for several pressures
  evidencing the quantum oscillations. It includes Refs.\
  \cite{liu2017superconductivity,qi2016pressure}}\ }\BibitemShut {NoStop}%
\bibitem [{EDM()}]{EDMOND}%
  \BibitemOpen
\bibfield  {journal} {  }\bibfield  {title} {\bibinfo {title} {See}\ }\href
  {https://doi.org/10.17617/3.P3ZAB1} {10.17617/3.P3ZAB1}\BibitemShut {NoStop}%
\end{thebibliography}%

	\begin{figure*}[!t]
	
	\centering
	\includegraphics[width=1\textwidth]{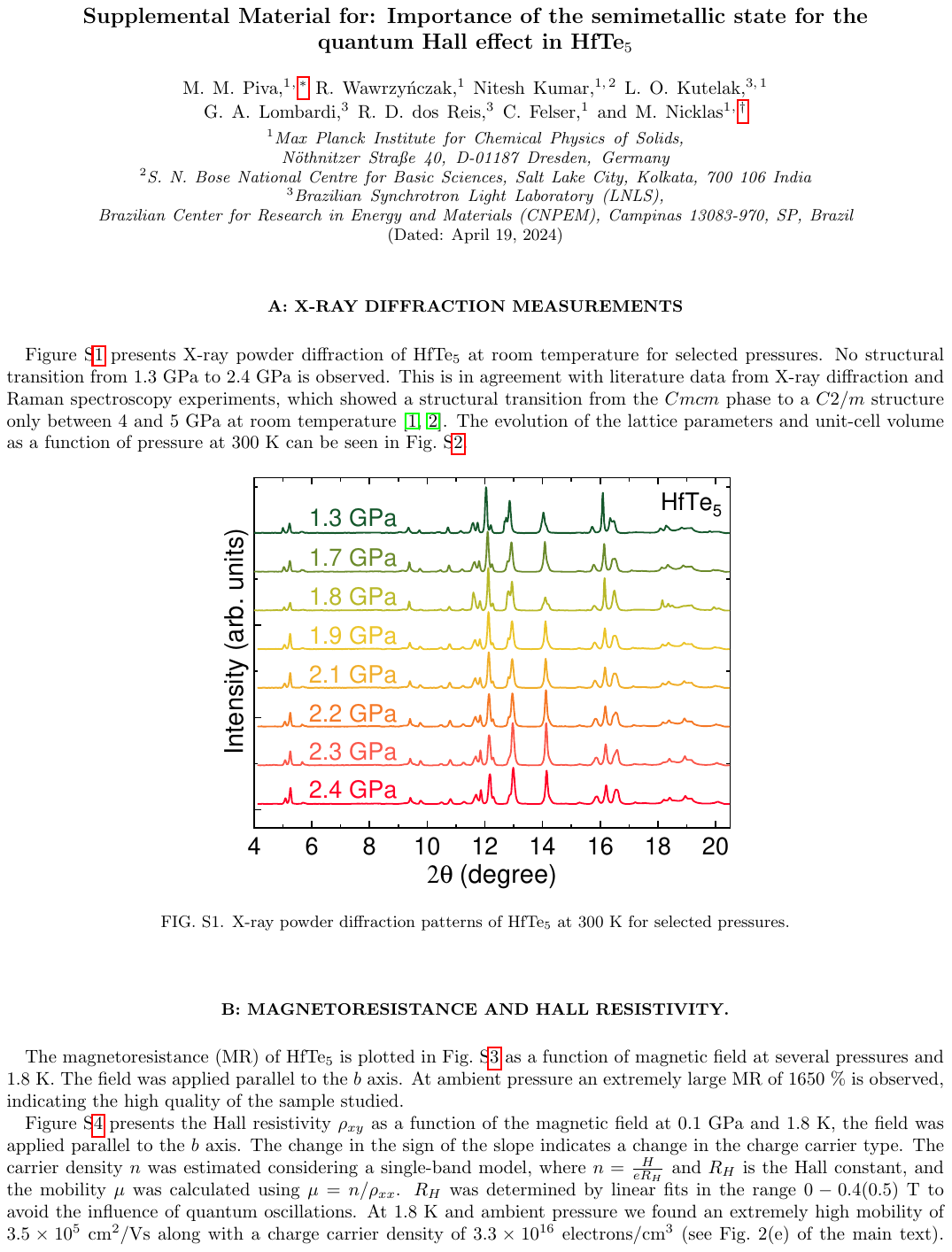}
\end{figure*}

	\begin{figure*}[!t]
	
	\centering
	\includegraphics[width=1\textwidth]{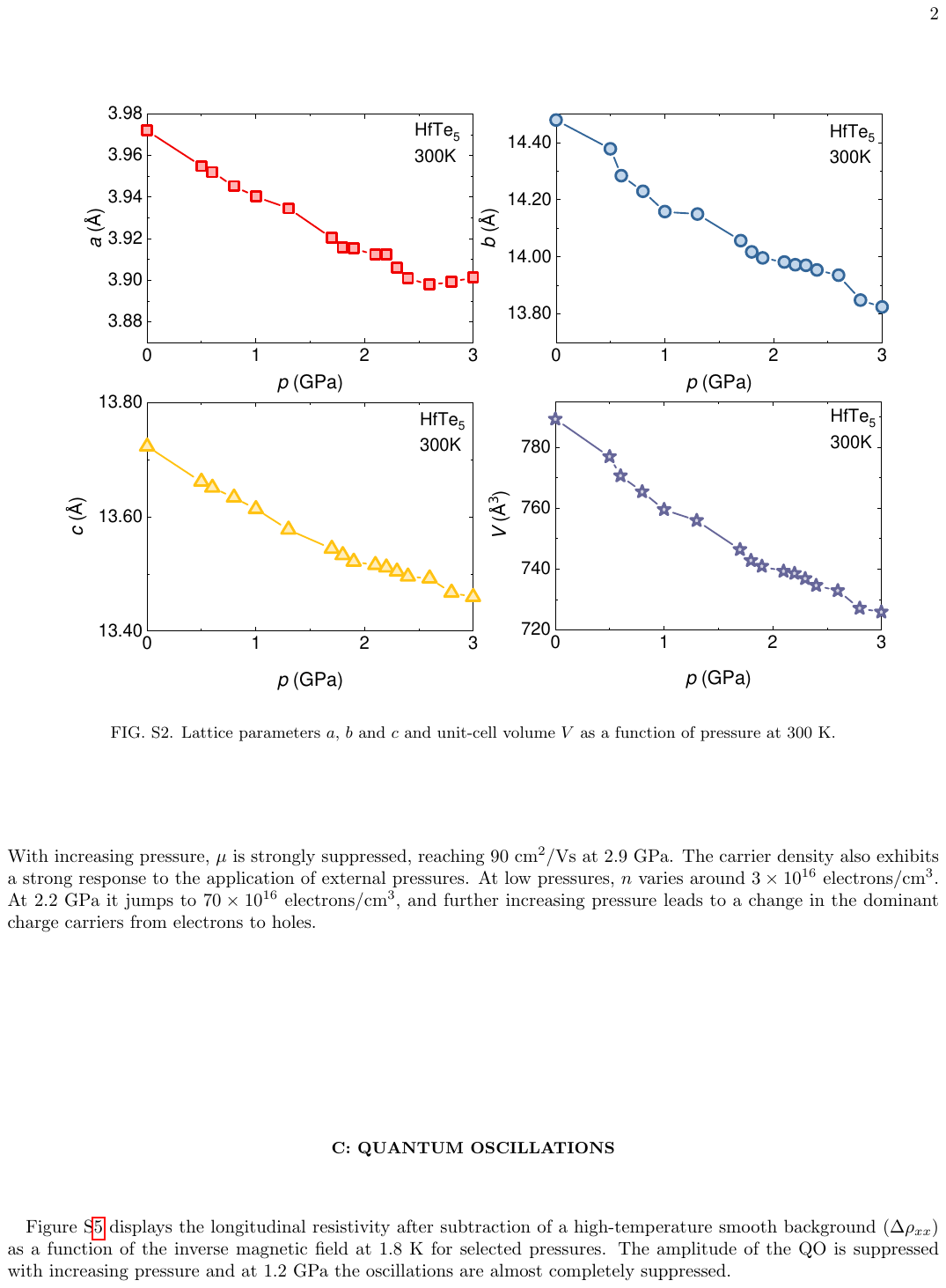}
\end{figure*}

	\begin{figure*}[!t]
	
	\centering
	\includegraphics[width=1\textwidth]{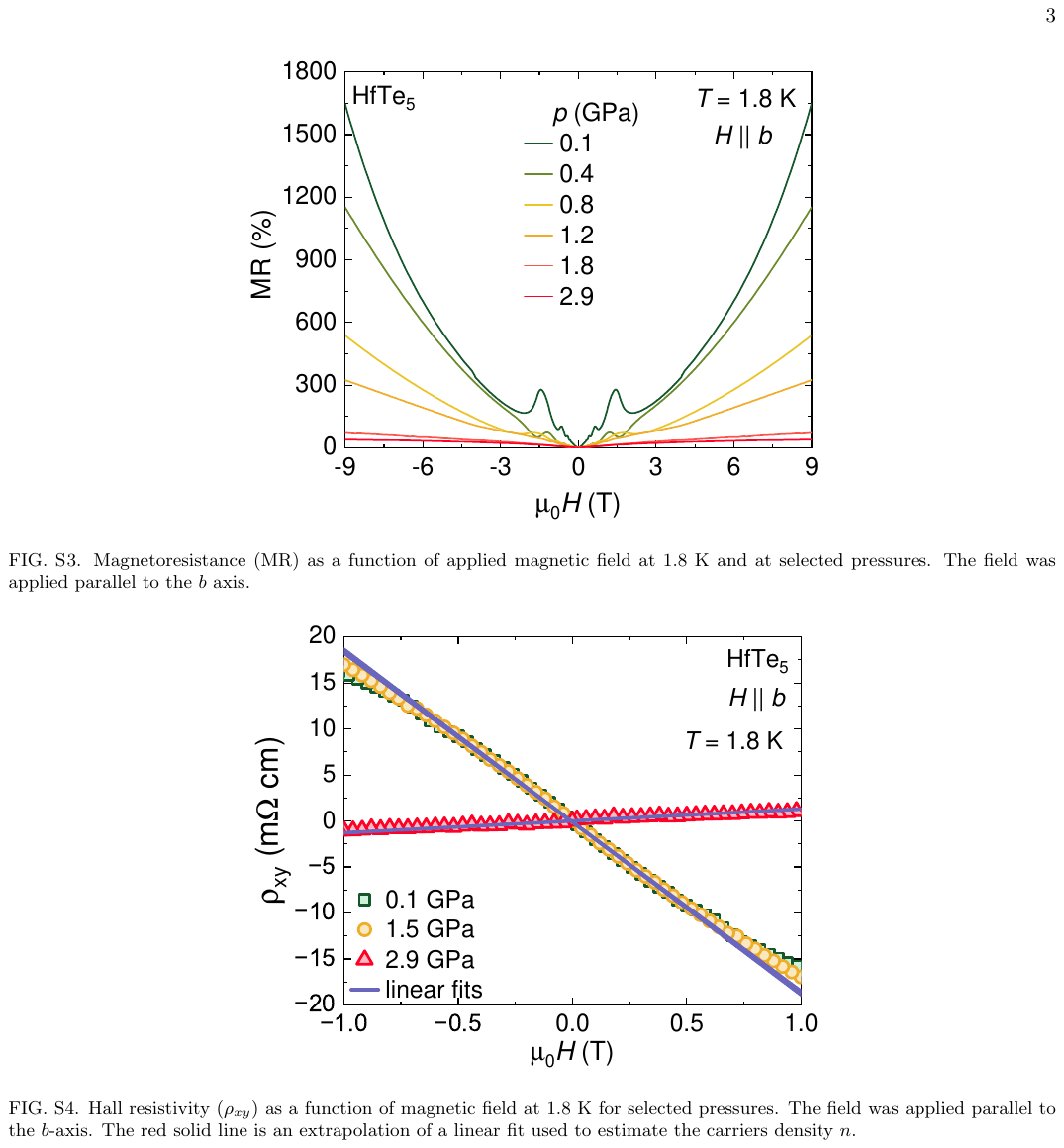}
\end{figure*}

	\begin{figure*}[!t]
	
	\centering
	\includegraphics[width=1\textwidth]{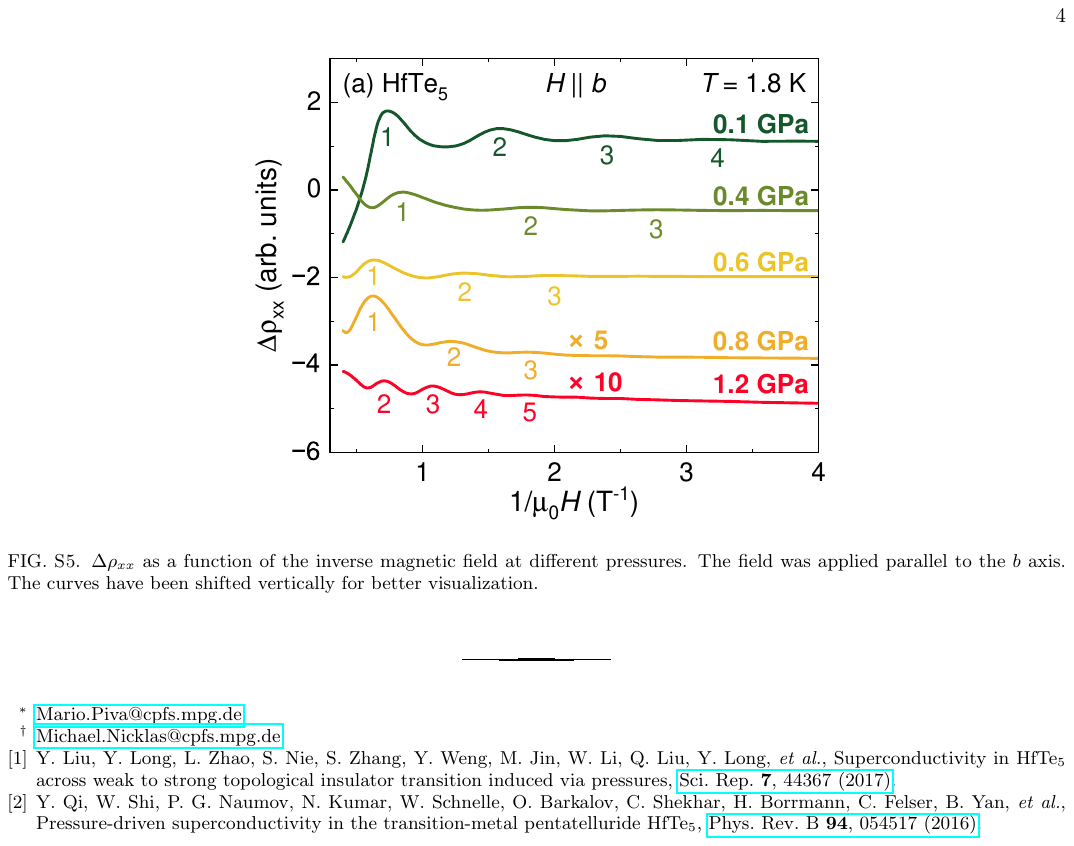}
\end{figure*}

\clearpage

\end{document}